\documentclass[a4paper,11pt]{article}
\pdfoutput=1 
\usepackage{jheppub} 
\title{\boldmath The BTZ black hole spectrum and partition function}
\author[a]{Omkar~Vinayak~Nippanikar,}
\author[a]{Aditya~Sharma,}
\author[b]{K.~P.~Yogendran.}
\affiliation[a]{Indian Institute of Science Education and Research Tirupati,\\Karakambadi Road, Mangalam, Tirupati, Andhra Pradesh, India}
\affiliation[b]{Indian Institute of Science Education and Research Mohali,\\Sector 81, Knowledge City, Mohali, Punjab, India}
\emailAdd{omkar.nippanikar@gmail.com}
\emailAdd{adsharma.d1d4@gmail.com}
\emailAdd{pattag@gmail.com}
\abstract{
In this article, we revisit the spectrum of the Lorentzian BTZ black hole conformal field theory. Building on a detailed analysis of geodesics, we identify a complete set of states for the harmonic analysis. We then demonstrate that the CFT spectrum is, plausibly, that of $AdS_3$ by rewriting the partition function of the latter CFT.
}


\arxivnumber{2112.11253}

\usepackage{wrapfig}
\usepackage{mathtools}
\usepackage[inline]{enumitem}
\usepackage{tikz}
\usetikzlibrary{decorations.pathmorphing}

\newcommand{\trace}{\text{Tr}}
\newcommand{\SL}[1]{$SL(#1,\mathbb{R})$}
\newcommand{\be}{\begin{equation}}
\newcommand{\ee}{\end{equation}}
\newcommand{\del}{\partial}
\newcommand{\e}{\epsilon}
\newcommand{\f}{\phi}
\newcommand{\caron}{\v}
\newcommand{\g}{\gamma}
\newcommand{\m}{\mu}
\newcommand{\p}{\pi}
\newcommand{\q}{\theta}
\newcommand{\s}{\sigma}
\newcommand{\z}{\zeta}
\newcommand{\D}{\Delta}
\newcommand{\F}{\Phi}
\newcommand{\G}{\Gamma}

\newcommand{\oao}[2]{{\biggl[\begin{array}{c}#1\\#2\end{array}\biggr]}} 
\newcommand{\ket}[1]{\mathchoice{\left|#1\right\rangle}{|#1\rangle}{|#1\rangle}{|#1\rangle}}

\begin{document} 
\maketitle
\flushbottom
\section{Introduction}\label{sec:Intro}
The BTZ black hole is a solution of 3D pure gravity with a negative cosmological constant. It also arises in string theory as a near-horizon region of a non-extremal system of 1- and 5- branes. This geometry has no curvature singularity because~\cite{btz} it is a discrete quotient of \SL2. The work of~\cite{mog, mog2} provided a complete understanding of the spectrum of the $AdS_3$ (\SL2) string sigma model. In particular, the authors argued that additional spectrally flowed representations of the \SL2 Ka\caron{c}-Moody algebra form a necessary part of the string spectrum. These representations were necessary to lift the bound on the mass of string states in $AdS_3$, otherwise imposed by the absence of ghosts. The `long strings' were then necessary to obviate the restriction on spacetime energy of spectrally flowed discrete series representations of the \SL2 conformal field theory (CFT) (while keeping the internal conformal weights fixed).

In~\cite{kpy}, the spectrum of the 2D Lorentzian black hole, which is a coset CFT of \SL2, was studied. In this case, even point-like states of the string theory were obtained after spectral flow with the spectral flow parameter being identified with the spacetime energy. The `spacelike' long strings appeared when the energy increased above the mass parameter of the state. The spectrum so constructed was argued to be consistent with an earlier proposal~\cite{Israel} for the partition function of this Lorentzian black hole. 

If we adopt the view that the black hole is a highly excited {\em state} (or density matrix) of the {\em global} $AdS_3$ string theory, then all the representations of the latter carry over as states of the black hole geometry. However, since the time coordinates of the two spacetimes do not obviously ``rotate" into each other, this continuation is not immediate. In fact, the twisted sectors of the BTZ orbifold arise from spectral flow along the orbifolding direction, which is different from the direction of spectral flow in the $AdS_3$ theory. Thus, it is not obvious if the twisted sector states of the BTZ black hole can be viewed as being states of the spectrally flowed sectors of the $AdS_3$ theory.

We know that in the case of $AdS_3$, the principal continuous series arose by quantising spacelike geodesics. In this case also, we expect to see the same, but there is a question of representation-theoretic interest. The principal continuous representations of \SL2, when written in the basis~\cite{Mukunda} that diagonalizes the `hyperbolic' generator $J^{(2)}$, contain two eigenvectors $\ket{\lambda,\pm}$ for each real eigenvalue $\lambda$ of $J^{(2)}$. In the 2D black hole~\cite{kpy}, the need for this doubling was associated with the presence of two distinct asymptotic regions. We can therefore try to understand the need for doubling of the spectrum in terms intrinsic to the BTZ geometry.

Finally, the extended BTZ geometry has regions containing closed timelike curves (CTCs). An important puzzle that is not yet resolved is how to handle these regions in string theory. In~\cite{NS}, it was suggested that these regions maybe be fully excised by using involutions in \SL2. However, the question of consistency can be considered settled only if one can exhibit a modular invariant partition function.  
We can expect that analysing the geodesics from the point of view of the WZW model could shed some light on these questions.

This document is organised as follows. After a brief recap of the geometry of the BTZ black hole as a discrete quotient of the \SL2 manifold, we present a detailed study of the geodesics of the black hole. In particular, we use this to identify a geodesically complete spacetime. We then examine the effect of certain discrete symmetries to understand the multiplicities of representations upon quantisation. We follow this up with a study of twisted sectors and the Virasoro conditions, which give us information about the possible states that make up the spectrum of this CFT. We then propose that the $AdS_3$ partition function of~\cite{Israel} contains all these identified states in a manner that is consistent with the orbifold CFT. Finally, we conclude with a summary and discussion. Special cases of the non-rotating and extremal black holes are discussed in Appendix~\ref{sec:SpecialCases}. Other appendices also contain supplementary material for a self-contained presentation.

\section{BTZ as an orbifold} \label{sec:BTZOrbifold}
We recap results from~\cite{btz,NS,kraus} about the BTZ geometry which is {\em defined} as the spacetime with the metric
\be
ds^2=-\frac{(r^2-r_+ ^2)(r^2-r_-^2)}{\ell^2r^2} dt^2+
\frac{\ell^2r^2}{(r^2-r_+ ^2)(r^2-r_-^2)}dr^2+r^2\left(d\f-\frac{r_+ r_-}{\ell r^2}dt\right)^2\,,\label{eqn:btzmetric}
\ee
where the coordinate $\f$ is periodic with period $2\p$. Henceforth, we set the $AdS_3$ radius $\ell=1$. If needed, this may be restored by the transformation $(ds,r,r_\pm,t)\mapsto\ell(ds,r,r_\pm,t)$. The metric above suggests that the coordinate $r^2$ has the range $[0,\infty)$. However, as we will see, most geodesics freely pass to regions beyond. Thus, we are forced to include additional regions to define a geodesically and causally complete manifold. The surface $r^2=0$ is termed the BTZ singularity even though the entire space has constant (negative) curvature, because the periodic $\phi$ coordinate becomes timelike in the region $r^2<0$. Hence this region contains CTCs.

This spacetime may also be viewed as (a part of) a discrete quotient of the \SL2 group manifold, which is 
defined as the set of real $2\times2$ matrices with unit determinant:
\be
g=\begin{pmatrix}x_1-x_2 & x_3-x_0\\
x_3+x_0  & x_1+x_2
\end{pmatrix}\,,\qquad\qquad\det g=1\,.
\ee
The quotienting relation is
\be
g \sim P_+ \, g\, P_- \qquad \forall \ g\in SL(2,\mathbb{R})\,, \qquad  P_\pm=e^{\p(r_ +\mp r_-)\s^3}\,.\label{eqn:orbifolding}
\ee 
The metric \eqref{eqn:btzmetric} is equal to the invariant metric (Killing-Cartan form) of \SL2 
provided we choose coordinates as described below. The special case of the extremal black hole with $r_+=r_-$ requires other forms for the charts~\cite{malstrom} that we shall discuss in Appendix~\ref{sec:SpecialCases}.

The \SL2 group manifold may be fully covered once, by matrices of the form
\be
g=(-\mathbb{I})^{\e_1}\,e^{\frac{1}{2}(r_+-r_-)(t+\f)\s^3}\,(i\s^2)^{\e_2}\,{\bf p}\,e^{-\frac{1}{2}(r_++r_-)(t-\f)\s^3}\,,\label{eqn:sl2charts}
\ee
where $\e_{1,2}\in\{0,1\}$ and ${\bf p}$ is one of $e^{\rho\s^1}$, $e^{i\rho\s^2}$ or $\left(\begin{smallmatrix}1&0\\\rho&1\end{smallmatrix}\right)$. In the first and last cases $\rho\in\mathbb{R}$ while in the second case $\rho\in[-\frac{\pi}{4},\frac{\pi}{4}]$. The third choice for ${\bf p}$ is applicable for \SL2 matrices with zero entries. It parametrises the horizons $r^2=r_\pm^2$. The BTZ identification \eqref{eqn:orbifolding} implies that the $\f$-coordinate is $2\pi$-periodic. Hence, the $(t,\f)$ coordinates define the ``boundary" cylinder at $\rho\to\infty$. 

Following~\cite{NS}, we shall refer to these coordinate charts using the notation $\pm D_i^\pm$ where $i\in\{1,2,3,4\}$, which is to be understood as follows. The charts $D_{1,4}$ use ${\bf p}=e^{\rho\s^1}$ while $D_{2,3}$ use ${\bf p}=e^{i\rho\s^2}$. The $D_{3,4}$ charts are obtained by setting $\e_2=1$ while the $D_{1,2}$ charts have $\e_2=0$. The sign in front of the $D_i$ represents $(-\mathbb{I})^{\e_1}$ while the sign in the superscript denotes the sign of $\rho$. When written explicitly, the chart $+D_1^+$ of \SL2 has matrices of the form
\be
g=e^{\frac{1}{2}(r_+-r_-)(t+\f)\s^3}
\begin{pmatrix}
\cosh\rho & \sinh\rho\\
\sinh\rho & \cosh\rho
\end{pmatrix}e^{-\frac{1}{2}(r_++r_-)(t-\f)\s^3}\,.\label{eqn:param}
\ee
With this parametrisation and the substitution $\cosh^2\rho=\frac{r^2-r_-^2}{r_+^2-r_-^2}$, the Killing-Cartan form $ds^2=-\frac{1}{2}\trace(dg^{-1}\,dg)$ becomes the BTZ metric \eqref{eqn:btzmetric}. Hence, the $+D_1^+$ chart covers the region $r^2>r_+^2$ outside the outer horizon.

We note here that global $AdS_3$ is a {\em cover} of \SL2 obtained by unwrapping the timelike elliptic (compact) direction in \SL2. This is, however {\em not} the time direction of the BTZ black hole.

The other \SL2 charts can also be mapped to other regions of the BTZ geometry. The region $r_-^2<r^2<r_+^2$ between the inner and outer horizons is parametrised by the $D_{2,3}$ charts with $\cos^2\rho=\frac{r^2-r_-^2}{r_+^2-r_-^2}$. The $D_4$ charts with $\sinh^2\rho=-\frac{r^2-r_-^2}{r_+ ^2-r_-^2}$ parametrise the region $r^2<r_-^2$ inside inner horizon. These charts also include the $r^2<0$ regions containing CTCs. Thus, the entire BTZ geometry (i.e., for all $r^2\in\mathbb{R}$) is fully covered once, by the atlas $\Omega_1=D_1 ^+\cup D_2 ^-\cup(-D_3 ^+)\cup(-D_4^-)$. This combination of charts ensures that the vector field $\frac{\del}{\del\rho}$ is continuous. The atlas $\Omega_2=D_1^-\cup D_2^+\cup D_3^-\cup D_4^+$ is defined to consist of complementary charts obtained from $\Omega_1$ by $\s^3$ conjugation. Together with two more atlases $-\Omega_{1,2}$, a single cover of the \SL2 group manifold includes four copies of the BTZ black hole.\footnote{The way these atlases cover \SL2 is described in Section~\ref{sec:GeometryOfSolutions}.} On the other hand, if we consider global $AdS_3$ which is the universal cover of \SL2 and then impose the BTZ identification, we get a geometry with infinitely many copies of the black hole and boundary components~\cite{kraus}.

\section{Geodesics of the BTZ}
The motion of a string on the BTZ geometry is governed by the action
\be
S_{\text{WZW}}[g]=-\frac{k}{8\pi} \int d\tau\,d\sigma\, \trace\left(\del_a gg^{-1}\del^a gg^{-1}\right)
+ k \Gamma_{\text{WZ}}[g]\,.\label{eqn:geodesicaction}
\ee
The Wess-Zumino term $\Gamma_{\text{WZ}}$ represents the pullback of the NS-NS B-field to the world volume of the string. Due to the WZ term, it can be seen that the right and left currents, defined as
\be
J_-=kg^{-1}\del_-g\,,\qquad\qquad J_+ = - k\del_+g g^{-1}
\ee
respectively, are chirally conserved i.e., $\del_\pm J_\mp=0$. Here, $x^\pm=\tau\pm\s$ and $\del_\pm=\frac12(\del_\tau\pm\del_\s)$. The worldsheet stress tensor of this theory has components
\be
T_{\pm\pm}=\frac{1}{2k}\trace(J_\pm J_\pm)\,.
\ee
In the parametrisations \eqref{eqn:sl2charts}, the action \eqref{eqn:geodesicaction} is invariant under $t$ and $\phi$ translations. Let us denote the associated conserved quantities as $E$ and $L$ respectively. These will turn out to be the energy and angular momentum from the point of view of asymptotic observer outside the outer horizon ($r^2>r_+^2$). When evaluated along geodesics (i.e., `collapsed' strings) these are expressible in terms of the currents as
\be
kE=-\D_-J^{(2)}_+-\D_+J^{(2)}_-\,,\qquad\qquad kL=-\D_-J^{(2)}_++\D_+J^{(2)}_-\,.\label{eqn:Energy}
\ee
Here, $\D_\pm=r_+\pm r_-$ and the components are computed as $J^{(2)}_\pm=\frac12\trace(\s^3 J_\pm)$. However, for stringy configurations (extended worldsheets), an additional total derivative term is required in the definition of the angular momentum for consistency with the level matching condition~\cite{NS,Esko,RR,GPS}.

The Euler Lagrange equations of motion for point-like trajectories gives us the geodesic equations on the BTZ geometry \eqref{eqn:btzmetric}. These equations~\cite{Cruz, Troost1} are
\begin{subequations}\label{eqn:geodesic}
\begin{align}
\dot y(\tau)^2&=4m^2\left(-y^2+\frac{\alpha}{m^2} y+\frac{\beta}{m^2}\right)\,,\label{eqn:geodesic:radial}\\
\dot t+\dot \f&=\frac{1}{(r_+-r_-)} \left(\frac{E r_+ +L r_-}{r_+^2-r ^2}+\frac{E r_- + L r_+}{r^2-r_- ^2}\right)\,,\label{eqn:geodesic:sum}\\
\dot t-\dot \f&=\frac{1}{(r_++r_-)}\left(\frac{E r_+ + L r_-}{r_+^2-r^2}-\frac{E r_- + L r_+}{r^2-r_-^2}\right)\,,\label{eqn:geodesic:difference}
\end{align}
\end{subequations}
where $y=r^2$, $\alpha=E^2-L^2+Mm^2$ and
$\beta=L^2M-\frac{1}{4}m^2J^2+ELJ$. It is perhaps worth mentioning a pedagogical point. The radial equation \eqref{eqn:geodesic:radial} has solutions with constant $y$ at roots of the quadratic on the right-hand side. However, these do not solve the second-order geodesic equations unless $\dot t+\dot\f$ or $\dot t-\dot\f$ vanish.

The equations of the WZW model amount to conservation of the left and right currents. For pointlike trajectories, these imply the geodesic equations. That is to say, the geodesic equations can be written as
\be
\del_\tau^2 g- \del_\tau g g^{-1}\del_\tau g=0 \quad\Longleftrightarrow\quad \del_\tau(g^{-1}\del_\tau g)=0=\del_\tau(\del_\tau gg^{-1})\,.
\ee
These equations are solved by any one parameter subgroup i.e., $g(\tau)=A\,\exp(m\tau T) B$ is a solution for any matrices $A,B\in$ \SL2 and where $T$ is an element of the Lie algebra of \SL2. Thus, we can rewrite the solutions as \SL2 matrices using the parametrisations \eqref{eqn:sl2charts}. In the process of uplifting the geodesics as \SL2 matrices, several sign choices have to be made. These choices can be interpreted as representing the same BTZ geodesics in different \SL2 regions/charts. Due to the symmetries of the BTZ geometry, one can always shift $t$ and $\f$ in the solution, which is equivalent to multiplying the solution matrix by constant diagonal \SL2 matrices on the left and right. 

We will now discuss the various solutions to the geodesic equations. We also focus on conditions that isolate geodesics which do not trespass into the regions containing CTCs.

\subsection{Timelike geodesics}
In this case, the solution to \eqref{eqn:geodesic:radial} is given by
\be
r^2(\tau)=\frac{\alpha}{2m^2}+\frac{\g}{2m^2}\sin (2m(\tau-\tau_0)+c)\,,\label{eqn:timelikeradial}
\ee
where we have defined $\alpha=E^2-L^2+m^2M$, $\beta=L^2M+ELJ-\frac{1}{4}m^2J^2$ and 
$\g^2=\alpha^2+4m^2 \beta$. Without loss of generality, we will set the integration constant $c=2m\tau_0$ to simplify various expressions, in what follows.

These trajectories, in general, oscillate from outside the outer horizon to inside the inner horizon. However, if $\beta<0$,  the trajectories do not pass through the singularity ($r^2=0$) into the region containing closed timelike curves. If $\alpha<0$, the trajectories already start behind the singularity at $\tau=0$. Imposing $\beta<0$ then ensures that these are prevented from entering the region $r^2>0$.

Using the radial solution \eqref{eqn:timelikeradial}, we can integrate \eqref{eqn:geodesic:difference} and \eqref{eqn:geodesic:sum}. The resulting solutions can lifted as \SL2 matrices using \eqref{eqn:param}. The lifts of timelike geodesics are
\be
\left(\frac{\g}{m^2(r_+^2-r_-^2)}\right)^{1/2}
\begin{pmatrix}
\frac{\sin(m\tau+\frac{\f_-}{2})} {\sqrt{\tan\frac{\f_-}{2}}} &
\frac{\cos(m\tau-\frac{\f_+}{2})}{\sqrt{\cot\frac{\f_+}{2}}} \\
\frac{\sin(m\tau+\frac{\f_+}{2})}{\sqrt{\tan\frac{\f_+}{2}}} & 
\frac{\cos(m\tau-\frac{\f_-}{2})}{\sqrt{\cot\frac{\f_-}{2}}}
\end{pmatrix}\,,
\label{eqn:timelikematrix}
\ee
where the angles $\f_\pm\in[0,\pi]$ are defined by $\cot \f_\pm=\frac { 2m(E r_\pm + L r_\mp)} {(E^2-L^2)\mp(r_+ ^2-r_-^2)m^2}$. Note that the normalisation factor written here is proportional to $\sqrt{\sin\f_--\sin\f_+}$. In particular, it may be purely imaginary. In these cases, the matrix is to be interpreted only after dropping the overall $i$ and flipping the sign of exactly one row or column (to retain $\det=1$).

The conserved quantities evaluated along the solution \eqref{eqn:timelikematrix} are
\be
T_{\pm\pm}=-\frac{k}{4}m^2\,,\qquad\qquad J^{(2)}_\pm=\pm\frac{k}{2}m\left[\cot\left(\frac{\phi_+\mp\phi_-}{2}\right)\right]^{\pm1}=-\frac{k (E\pm L)}{2(r_+ \mp r_-)}\,.\label{eqn:timelikecurrents}
\ee
Since $T_{\pm\pm}<0$, upon quantisation, we can expect that these will belong to the discrete series representation of \SL2.

From the definition of $\cot\f_\pm$, we see that special solutions occur when $\f_\pm=0$ or $\f_\pm=\frac{\pi}{2}$. The condition $\f_\pm=0$ is equivalent to $E^2-L^2=\pm m^2(r_+ ^2-r_- ^2)$ and can be rewritten as
\be
J^{(2)}_+ J^{(2)}_-=\mp kT_{\pm\pm}\,.
\ee
These solutions satisfy a relation
\be
\cos[2(r_\mp t-r_\pm\f)]=\frac{1}{2}\left(\frac{r_+^2-r_-^2}{r^2-r_\pm^2}\right)\,.
\ee
The second set of special solutions which satisfy $\f_\pm=\frac{\pi}{2}$ obey
\be
J^{(2)}_+=\mp J^{(2)}_- \quad\Longleftrightarrow\quad r_\pm t-r_\mp\f={\text{constant}}\,.\label{eqn:TimelikeGrazingHorizons}
\ee
These geodesics touch the horizon $r^2=r_\pm^2$ before turning back. Clearly, the geodesics with $\f_-=\frac{\p}{2}$ (and arbitrary $\f_+$) do not explore the region with CTCs and indeed satisfy $\beta<0$. In the further special case $\f_+=\frac{\p}{2}=\f_-$, they also remain between the horizons. The other set $\f_+=\frac{\p}{2}$ do not necessarily satisfy $\beta<0.$

Each time the geodesic \eqref{eqn:timelikematrix} crosses a horizon, one of the matrix entries becomes zero (see chart \eqref{eqn:sl2charts}).  
Thus, there are three natural time scales that appear in the solution \eqref{eqn:timelikematrix} viz., the proper time taken to traverse: \begin{enumerate*}[label=(\alph*)]\item from the past outer horizon to the future outer horizon $\D\tau_+$; \item from the past inner horizon to the future inner horizon $\D\tau_-$; \item between the outer and inner horizons $\D\tau_0$ \end{enumerate*}. 

The proper time interval to traverse from past horizon to future horizon (while outside the outer horizon) is $|m\D\tau_+|=|\f_+-\frac{\pi}{2}|$. Similarly, the proper time taken to traverse from past horizon to future horizon (while inside the inner horizon) is $|m\D\tau_-|=|\f_--\frac{\pi}{2}|$. Also, the proper time taken to traverse from outer horizon to inner horizon (while between the horizons) is the lesser of $|m\D\tau_1|=|\frac{\pi}{2}-\frac{1}{2}(\f_++\f_-)|$ and $|m\D\tau_2|=\frac{1}{2}|\f_+-\f_-|$. Remarkably, all these time-scales can be expressed in terms of the currents as
\begin{subequations}\label{eqn:timeliketimescale}
\begin{gather}
\tan |m\D\tau_\pm|=\left|\frac{km}{2}\left(\frac{J^{(2)}_+\pm J^{(2)}_-}{J_+^{(2)}J_-^{(2)}\mp kT}\right)\right|\,,\\
\tan |m\D\tau_1|=\left|\frac{km}{2J^{(2)}_-}\right|\,,\qquad\qquad\tan |m\D\tau_2|=\left|\frac{km}{2J^{(2)}_+}\right|\,,
\end{gather}
\end{subequations}
where $T=T_{\pm\pm}=-\frac{k}{4}m^2$ denotes the stress tensor evaluated along the timelike geodesic \eqref{eqn:timelikematrix}. The fact that all these timescales are well defined is consistent with the fact that the timelike geodesics cross all four horizons at finite proper times. 

\subsection{Spacelike geodesics}
As in $AdS_3$, we see that none of the timelike geodesics reaches the boundary observer. We may expect spacelike geodesics to include such scattering solutions.

Spacelike geodesics of the BTZ black hole were also considered in~\cite{Troost1}. We write the solutions as
\be
r^2(\tau)=\frac{-\alpha}{2m^2}\pm\frac14\left(e^{2m\tau}+\frac{\alpha^2-4m^2\beta}{m^4}e^{-2m\tau}\right)\,,\label{eqn:spacelikegeneral}
\ee
where now $\alpha=E^2-L^2-m^2M$ and $\beta=L^2M+ELJ+\frac14 m^2J^2$. If $\alpha^2-4m^2\beta\neq0$ then, this may be rewritten upto a shift of $\tau$ as
\begin{subequations}\label{eqn:spacelikeradial}
\begin{align}
r^2(\tau)&=\frac{1}{2m^2}\left(-\alpha\pm\sqrt{\alpha^2-4m^2\beta}\cosh2m\tau\right) &{\text{if }}\alpha^2-4m^2\beta>0\,,\label{eqn:spacelikeradial:positive}\\
r^2(\tau)&=\frac{1}{2m^2}\left(-\alpha\pm\sqrt{|\alpha^2-4m^2\beta|}\sinh2m\tau\right) &{\text{if }}\alpha^2-4m^2\beta<0\,.\label{eqn:spacelikeradial:negative}
\end{align}
\end{subequations}
First, we observe that the geodesics \eqref{eqn:spacelikeradial:negative} with $\alpha^2-4m^2\beta < 0$ necessarily cross into the region containing CTCs. For \eqref{eqn:spacelikeradial:positive} with the upper sign choice, there is a minimum distance of approach $r_*^2=\frac{-\alpha}{2m^2}+\frac{1}{2m^2}\sqrt{\alpha^2-4m^2\beta}$. This is positive if $\alpha<0$ or $\beta<0$, whence the geodesic stays out of the region with CTCs. Unlike timelike geodesics, there are solutions that stay out of this region but do not satisfy $\beta<0$. Interestingly, the constraint $\beta<0$ does not allow for values $|E|<\sqrt{m^2M}$ of the energy irrespective of the angular momentum $L$. However, these values of energy are allowed if $\alpha<0$ instead.

With the angular equations \eqref{eqn:geodesic:sum} and \eqref{eqn:geodesic:difference} integrated, we can uplift the solutions \eqref{eqn:spacelikeradial:positive} as the \SL2 matrices
\be
\begin{split}
\frac{\begin{pmatrix}
e^{-m\tau}A_{11}+e^{m\tau}& e^{-m\tau}A_{12}+e^{m\tau}\\
e^{-m\tau}A_{21}+e^{m\tau}& e^{-m\tau}A_{22}+e^{m\tau}
\end{pmatrix}}{\sqrt{4m^2(r_+^2-r_-^2)}}\,,
\end{split}\qquad
\begin{split}
\begin{aligned}
A_{11}&=(L-mr_+)^2-(E+mr_-)^2\,,\\
A_{12}&=(L+mr_-)^2-(E-mr_+)^2\,,\\
A_{21}&=(L-mr_-)^2-(E+mr_+)^2\,,\\
A_{22}&=(L+mr_+)^2-(E-mr_-)^2\,.
\end{aligned}
\end{split}\label{eqn:spacelikematrix}  
\ee
The lift of \eqref{eqn:spacelikeradial:positive} with lower sign choice is obtained by flipping the signs of all $A_{ij}$ followed by sign flips of exactly one row or column. The worldsheet stress tensor and the conserved charges evaluated along this solution are
\be
T_{\pm\pm}=\frac{k}{4}m^2\,,\qquad\qquad J^{(2)}_\pm=-\frac{k(E\pm L)}{2(r_+\mp r_-)}\,.\label{eqn:spacelikecurrents}
\ee
Since $T_{\pm\pm}>0$, we can expect these geodesics to arise from the principal continuous series of \SL2 upon quantisation.

The lifted solution can also be written purely in terms of these conserved charges using $k^2A_{ij}=(r_-^2-r_+^2)(2J^{(2)}_+-(-1)^ikm)(2J^{(2)}_-+(-1)^jkm)$. Clearly, special solutions occur at $(J_\pm^{(2)})^2=kT_{\pm\pm}$ when $\alpha^2-4m^2\beta=0$. This implies that $r^2$ asymptotes to a constant as $\tau\to-\infty$. We can see that these solutions satisfy
\begin{subequations}
\begin{align}
e^{\pm2(r_++r_-)(t-\f)}&=\frac{r^2-r_-^2}{r^2-r_+^2}\,, &E+L&=\pm m(r_+-r_-)\,, &J^{(2)}_+&=\mp\frac{k}{2}m\,;\\
e^{\pm 2(r_+-r_-)(t+\f)}&=\frac{r^2-r_-^2}{r^2-r_+^2}\,, &E-L&=\pm m(r_++r_-)\,, &J^{(2)}_-&=\mp\frac{k}{2}m\,.
\end{align}
\end{subequations}

For another special case $J^{(2)}_+=\pm J^{(2)}_-$, the diagonal or off-diagonal entries become independent of $E$ and $L$ after a shift of $\tau$. These solutions satisfy $\alpha^2-4m^2\beta=(\alpha+2m^2r_\mp^2)^2$ and similar to the timelike case, these spacelike geodesics touch the horizons at $r^2=r_\mp^2$, as seen from \eqref{eqn:spacelikegeneral}.

As for the timelike geodesics, we can calculate the time scales associated to a spacelike geodesic with $\alpha^2-4m^2\beta>0$ (and upper sign choice in \eqref{eqn:spacelikeradial:positive}). As before, let us denote the proper time taken to traverse: \begin{enumerate*}[label=(\alph*)] \item from the past outer horizon to the future outer horizon as $\D\tau_+$; \item from the past inner horizon to the future inner horizon as $\D\tau_-$; \item from outer horizon to inner horizon as $\D\tau_0$ \end{enumerate*}. These time-scales can be expressed in terms of the currents \eqref{eqn:spacelikecurrents} as
\begin{subequations}\label{eqn:spaceliketimescale}
\begin{align}
\tanh |m\D\tau_\pm|&=\left|\frac{km}{2}\left(\frac{J^{(2)}_+\pm J^{(2)}_-}{J_+^{(2)}J_-^{(2)}\pm kT}\right)\right|\,,\\
\tanh |m\D\tau_0|&={\text{min}}\left\{\left|\frac{km}{2J_+^{(2)}}\right|,\left|\frac{km}{2J_-^{(2)}}\right|\right\}\,,
\end{align}
\end{subequations}
where $T=T_{\pm\pm}=\frac{k}{4}m^2$ denotes the stress tensor \eqref{eqn:spacelikecurrents} evaluated along the spacelike geodesic. It is to be noted that the right hand side needs to be less than unity for the corresponding time scale to be defined. This will be the case, whenever the geodesic is such that it crosses the appropriate horizons. We will return to this point in Section~\ref{sec:GeometryOfSolutions} where we discuss the geometry of the solutions. 

\subsection{Null geodesics}
The most general null geodesic of the BTZ black hole is
\begin{subequations}\label{eqn:nullradial}
\begin{align}
\alpha r^2(\tau)&=\left(\alpha\tau+\sqrt{\alpha r_0^2+\beta}\right)^2-\beta &{\text{if }}\alpha\neq0\,,\label{eqn:nullradial:nonzero}\\
r^2(\tau)&=2E(r_+\pm r_-)\tau+r_0^2 &{\text{if }}\alpha=0\,,
\end{align}
\end{subequations}
where $\alpha=E^2-L^2$, $\beta=L(LM+EJ)$ and $r_0=r(0)$. 
When $\alpha>0$, these geodesics reach the boundary of $AdS_3$. If $\alpha<0$, they inevitably penetrate through to the region $r^2<0$ which contains CTCs. If $\alpha>0$ they always penetrate inside the inner horizon because ${\text{min}}\{\alpha(r^2(\tau)-r_-^2)\}=-(Er_-+Lr_+)^2\leq0$. In addition to $\alpha>0$, if $\beta<0$ these geodesics do not pass into the region $r^2<0$ containing CTCs. This maybe viewed as a bound $\frac{|E|}{M}>|\frac{L}{J}|$ on the energy $E$ which is stronger than $E^2>L^2$ (since $|J|<M$).

The lifts of null geodesics to \SL2 (upto a $\tau$ shift) are
\begin{subequations}\label{eqn:nullmatrix}
\begin{align}
\frac{1}{\sqrt{|\alpha|(r_+ ^2-r_- ^2)}}
&\begin{pmatrix}
\alpha\tau-(Er_-+Lr_+) & \alpha\tau+(Er_++Lr_-)\\
\alpha\tau-(Er_++Lr_-) & \alpha\tau+(Er_-+Lr_+)
\end{pmatrix} & {\text{if }}\alpha>0\,,\label{eqn:nullmatrix:plus}\\
\frac{1}{\sqrt{|\alpha|(r_+ ^2-r_- ^2)}}
&\begin{pmatrix}
\alpha\tau-(Er_-+Lr_+) & -\alpha\tau-(Er_++Lr_-)\\
\alpha\tau-(Er_++Lr_-) & -\alpha\tau-(Er_-+Lr_+)
\end{pmatrix} & {\text{if }}\alpha<0\,.\label{eqn:nullmatrix:minus}
\end{align}
\end{subequations}
The geodesics \eqref{eqn:nullradial:nonzero} cross each of the inner and outer horizons exactly twice. 
As in the case of spacelike geodesics \eqref{eqn:spacelikeradial:positive}, null geodesics also reach the boundary at a finite time ($t=0$, in particular). For this reason, we expect that the null geodesics form a special case of spacelike geodesics. 

The conserved quantities evaluated along the solution \eqref{eqn:nullmatrix} are
\be
T_{\pm\pm}=0\,,\qquad\qquad J^{(2)}_\pm=-\frac{k(E\pm L)}{2(r_+\mp r_-)}\,.\label{eqn:nullcurrents}
\ee
The time scales associated to these geodesics
\begin{subequations}
\begin{align}
\D\tau_\pm&=\frac{k}{2}\left|\frac{J^{(2)}_+\pm J^{(2)}_-}{J^{(2)}_+J^{(2)}_-}\right|\,,\\
\D\tau_0&={\text{min}}\left\{\left|\frac{k}{2J^{(2)}_+}\right|,\left|\frac{k}{2J^{(2)}_-}\right|\right\}
\end{align}
\end{subequations}
are seen to be the $m\to0$ limits of corresponding time scales \eqref{eqn:spaceliketimescale} for spacelike geodesics as well as \eqref{eqn:timeliketimescale} for timelike geodesics.

To summarise, we see that for timelike geodesics, a single condition $\beta<0$ keeps them out of the region containing CTCs. On the other hand, there are spacelike geodesics which do not satisfy $\beta<0$ but nevertheless stay out of this region.

\section{Geometry of the solutions}\label{sec:GeometryOfSolutions}
\begin{wrapfigure}[41]{l}{0.3\linewidth}
\begin{tikzpicture}
\fill[gray!25] (4,0)--(0,4)--(4,8)--cycle;
\fill[lime!25] (0,4)--(4,8)--(0,12)--cycle;
\draw[black] (0,0)--(4,0);
\draw[black] (0,4)--(4,4);
\draw[black] (0,8)--(4,8);
\draw[black] (0,12)--(4,12);
\draw[black] (0,16)--(4,16);
\draw[black] (0,0)--(0,16);
\draw[black] (4,0)--(4,16);
\draw[blue,thick] (0,0)--(4,4);
\draw[blue,thick] (0,4)--(4,0);
\draw[blue,thick] (0,8)--(4,12);
\draw[blue,thick] (0,12)--(4,8);
\draw[teal,thick] (0,4)--(4,8);
\draw[teal,thick] (0,8)--(4,4);
\draw[teal,thick] (0,12)--(4,16);
\draw[teal,thick] (0,16)--(4,12);
\draw[red,thick,decorate,decoration={snake,amplitude=1pt,segment length=4pt}] (0,4) to [out=75, in=-75] (0,8);
\draw[red,thick,decorate,decoration={snake,amplitude=1pt,segment length=4pt}] (0,12) to [out=75, in=-75] (0,16);
\draw[red,thick,decorate,decoration={snake,amplitude=1pt,segment length=4pt}] (4,4) to [out=105, in=-105] (4,8);
\draw[red,thick,decorate,decoration={snake,amplitude=1pt,segment length=4pt}] (4,12) to [out=105, in=-105] (4,16);
\node at (3,2) [] {$+D_1^+$};
\node at (3,6) [] {$-D_4^-$};
\node at (3,10) [] {$-D_1^+$};
\node at (3,14) [] {$+D_4^-$};
\node at (1,2) [] {$+D_1^-$};
\node at (1,6) [] {$-D_4^+$};
\node at (1,10) [] {$-D_1^-$};
\node at (1,14) [] {$+D_4^+$};
\node at (2,1) [] {$+D_2^+$};
\node at (2,3) [] {$+D_2^-$};
\node at (2,5) [] {$-D_3^+$};
\node at (2,7) [] {$-D_3^-$};
\node at (2,9) [] {$-D_2^+$};
\node at (2,11) [] {$-D_2^-$};
\node at (2,13) [] {$+D_3^+$};
\node at (2,15) [] {$+D_3^-$};
\draw[blue,thick] (0.2,-0.75)--(0.7,-0.75);
\node at (1.25,-0.75) [] {$r^2=\mathrlap{r_+^2}$};
\draw[teal,thick](0.2,-1.25)--(0.7,-1.25);
\node at (1.25,-1.25) [] {$r^2=\mathrlap{r_-^2}$};
\draw[red,thick,decorate,decoration={snake,amplitude=1pt,segment length=4pt}] (0.2,-1.75)--(0.7,-1.75);
\node at (1.25,-1.75) [] {$r^2=\mathrlap{0}$};
\node[rectangle,draw=gray!100,fill=gray!25,very thick,minimum size=10pt] at (3.2,-0.8) {$+\Omega_1$};
\node[rectangle,draw=lime!100,fill=lime!25,very thick,minimum size=10pt] at (3.2,-1.6) {$-\Omega_2$};
\end{tikzpicture}
\caption{Penrose diagram of the single cover of \SL2 group manifold}\label{fig:Penrose}
\end{wrapfigure}
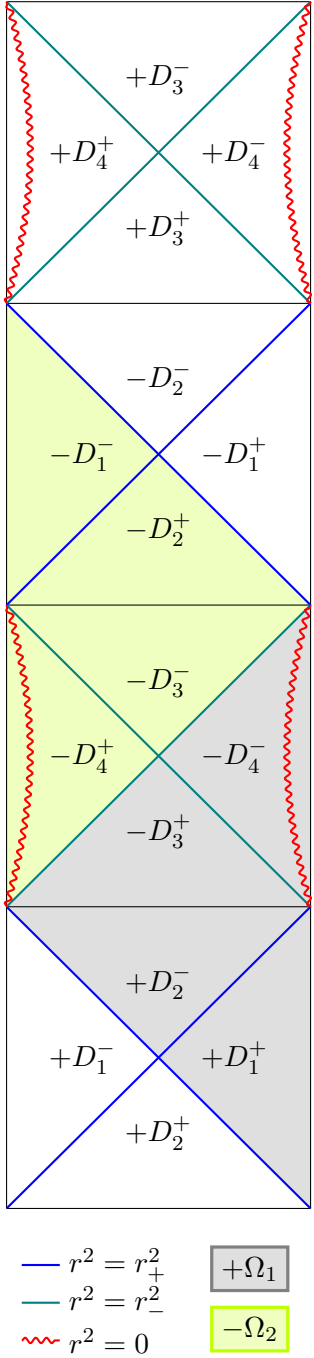
In this section, we shall pay close attention to the {\em trajectories} of geodesics as they traverse various regions of the BTZ geometry. Using the matrices representing each class of solutions, we can track their evolution through the Penrose diagram of \SL2. This shows the manner in which the various regions and copies of the BTZ geometry sit inside \SL2 (in this consideration, we ignore the orientation of the affine parameter $\tau$). An issue in the physics of the BTZ geometry is the presence of regions containing CTCs. Hence, the question of their consistent excision is important. A second issue is that four copies of the BTZ geometry cover \SL2 once. Therefore, we will attempt to gather a minimal set of charts from $\{+D_i^\pm,-D_i^\pm\}$ required to make a geodesically complete spacetime by first considering the null and timelike geodesics. Subsequently, we shall discuss spacelike geodesics since these are important only in the string theory.

Figure~\ref{fig:Penrose} shows the Penrose diagram of the \SL2 group manifold showing the regions covered by the various charts in the notation discussed in Section~\ref{sec:BTZOrbifold}. Consider the null geodesic \eqref{eqn:nullmatrix:plus} with $J^{(2)}_+>J^{(2)}_->0$. This geodesic for large negative $\tau$ has all entries of the matrix negative and thus lies in the chart $-D_1 ^+$. As $\tau$ increases, the number of negative entries changes gradually from four to none and the geodesic passes through the charts \[-D_1 ^+\to -D_2 ^+\to -D_3 ^- \to -D_4 ^- \to -D_3 ^+ \to +D_2 ^- \to +D_1 ^+\] finally ending in the chart $D_1^+$ for large positive $\tau$. Similarly, the order in which timelike geodesics traverse the charts can also be found. For instance, if $0<\f_+<\p-\f_-<\frac{\p}{2}\Leftrightarrow {\text{sgn}}(m)J^{(2)}_\pm<0$ then the timelike geodesic \eqref{eqn:timelikematrix} traverses the charts in the order \[+D_1^+\to+D_2^-\to-D_3^+\to-D_4^+\to-D_3^-\to-D_2^+\to-D_1^+\] as $m\tau$ varies in the range $[0,\p]$. Evolution beyond $m\tau\in[0,\p]$ may be found using the observation that $m\tau\mapsto m\tau+\p$ is equivalent to multiplication by $-\mathbb{I}$. This analysis can also be repeated for other values of $J^{(2)}_\pm$. Of particular interest are the special geodesics \eqref{eqn:TimelikeGrazingHorizons} with $\f_-=\frac{\p}{2}$ (and $\f_+=\frac{\p}{2}$). The matrices \eqref{eqn:timelikematrix} corresponding to them have equal diagonal (anti-diagonal) entries and hence, they vanish together at some value of $\tau$. Such geodesics pass through the intersection of the future and past horizons $r^2=r_- ^2$ (and $r^2=r_+ ^2$).

The spacelike geodesics however start and end in asymptotic regions. Here, we reproduce the matrix
\be
\sqrt{\frac{r_+^2-r_-^2}{4m^2}}
\left(\begin{smallmatrix}
e^{m\tau}-\frac{e^{-m\tau}}{k^2}\left\{(2J^{(2)}_--km)(2J^{(2)}_++km)\right\} & e^{m\tau}-\frac{e^{-m\tau}}{k^2}\left\{(2J^{(2)}_-+km)(2J^{(2)}_++km)\right\}\\
e^{m\tau}-\frac{e^{-m\tau}}{k^2}\left\{(2J^{(2)}_--km)(2J^{(2)}_+-km)\right\} & e^{m\tau}-\frac{e^{-m\tau}}{k^2}\left\{(2J^{(2)}_-+km)(2J^{(2)}_+-km)\right\}
\end{smallmatrix}\right)\,,\label{eqn:SpacelikeExample}
\ee
that represents the spacelike geodesic \eqref{eqn:spacelikegeneral} with upper sign choice (upto a $\tau$ shift). When $J^{(2)}_+>\frac{k|m|}{2}$ and $J^{(2)}_-<-\frac{k|m|}{2}$, terms in braces in the above matrix are all negative and none of the matrix entries change sign as $\tau$ varies. Hence, these geodesics start and end in $+D_1^+$. Note that although the right hand sides of \eqref{eqn:spaceliketimescale} are less than unity, the time scales $\D\tau_\pm$ and $\D\tau_0$ for this geodesic are not defined, because it never crosses any horizon. On the other hand, when $|J^{(2)}_\pm|<\frac{k|m|}{2}$, only the terms in braces on the diagonal are negative. When $J^{(2)}_-=0=J^{(2)}_+$, such geodesics pass through the bifurcation point from $+D_1^-$ to $+D_1^+$. More generally, they traverse via $+D_2^\pm$ if $\pm m(J^{(2)}_++J^{(2)}_-)<0$. Since they cross only the outer horizon twice, it is satisfactory that $\tanh|m\D\tau_0|>1$ in \eqref{eqn:spaceliketimescale}, thereby rendering the time scale $\D\tau_0$ undefined. However, although $\tanh|m\D\tau_\pm|<1$, only the time scale $\D\tau_+$ is applicable to these geodesics. These are clearly folded strings, but rather than ending at the horizon, these end on the boundary much as in the 2D black hole~\cite{kpy}.

There are also geodesics that end in $D_1^+$ starting from one of the $D_4$ charts. For instance, geodesics with $-km<2J^{(2)}_+<km$ and $-2J^{(2)}_-<km<2J^{(2)}_-$ start from $-D_4^+$ and end in $D_1^+$. Similar analyses can be repeated for spacelike geodesics \eqref{eqn:spacelikegeneral} with the lower sign choice.

From this analysis, it is clear that each of the $\{+D_i^\pm,-D_i^\pm\}$ charts can be accessed from every other chart by traversing (piecewise) geodesic trajectories. Hence, geodesic completeness requires that we keep all these charts. In particular, we are forced to include the $D_4$ charts containing the regions with CTCs. This will mean that the quantum wavefunctions corresponding to these geodesics will certainly have nonzero support in these charts.

\section{Discrete symmetries} \label{sec:DiscreteSymmetries}
In this section, we shall consider the action of various discrete symmetries to get a handle on the multiplicities of solutions with the same quantum numbers. Our approach will be in the spirit of covariant quantisation -- we will examine the effect of the symmetries on the {\em space} of solutions. Some of these discrete symmetries, such as time-reversal, are interpretable at the level of the BTZ geometry itself. Some arise naturally by considering the matrix representation of the solutions.

\subsection{BTZ symmetries}
The BTZ metric exhibits several discrete symmetries. It is invariant under a simultaneous sign flip of $t$ and $\f$. It is also invariant under $r_+\leftrightarrow r_-$.  Since these transformations preserve the geodesic {\em equations} \eqref{eqn:geodesic}, they act on a solution (a geodesic) and produce another solution, in general. For instance, the transformation $(t,\f)\mapsto(-t,-\f)$ when applied to a solution with energy $E$ and angular momentum $L$ gives us a solution with energy $-E$ and angular momentum $-L$. However, all these transformations leave the quantities $\alpha$ and $\beta$ invariant.

The string {\em sigma model} with the BTZ geometry as target space is also invariant under $r_+\leftrightarrow r_-$. This is because the NS B-field can be written, up to a gauge trivial term as
\be
B_{NS}=(r^2-r_+ ^2-r_-^2)\;d\f\wedge dt\,. \label{eqn:BTZSymmetryB-field}
\ee
This transformation is akin to quotienting by the `vectorial' action of $P_+$ in \eqref{eqn:orbifolding} if the earlier action is termed `axial'.

Interchanging $E\leftrightarrow L$ together with $t\leftrightarrow\f$ and $r^2-r_+ ^2\leftrightarrow r_-^2-r^2$ also leaves the geodesic equations invariant~\cite{NS}. To see this, it helps to rewrite the radial equation \eqref{eqn:geodesic:radial} as
\be
\frac{\dot y^2}{4m^2}= 
-\left(y-\frac{M}{2}\right)^2+\frac{E^2-L^2}{m^2}\left(y-\frac{M}{2}\right)+\frac{M(E^2+L^2)+2ELJ}{2m^2}+\frac{M^2-J^2}{4}\,.
\ee
The interchange $r^2-r_+ ^2\leftrightarrow r_-^2-r^2$ now corresponds to $y-\frac{M}{2}\mapsto \frac{M}{2}-y$. Along with $E\leftrightarrow L$ and $t\leftrightarrow\f$, this leaves all the geodesic equations \eqref{eqn:geodesic} invariant. This transformation can be implemented on any solution of the WZW model by right multiplying the solution matrix by $i\s_2$. When acting on a {\em geodesic} having energy and angular momentum $(E,L)$, it results in a new geodesic with energy and angular momentum $(E^\prime,L^\prime)=(L,E)$. In fact, the matrix obtained by right multiplying any solution of the {\em WZW model} by $i\s_2$ is also a solution. If we replace the worldsheet $\tau$ variable by $\s$, this solution will turn out to have same conserved charges as the original geodesic. This is analogous to usual T-duality where we obtain new states of the theory by $\tau\leftrightarrow\s$, thereby changing sign of the right moving current $J_-$. This procedure was used to obtain new stringy states of the 2D Lorentzian black hole~\cite{kpy}.

To summarise, we realised that the discrete symmetry $(t,r_+^2-r^2)\leftrightarrow(\f,r^2-r_-^2)$ of the BTZ sigma model acts on lifts of its solutions to \SL2 as right multiplication by $i\s^2$. Hence, it is part of the \SL2$\times$\SL2 symmetry of the WZW model and does not lead to multiplicities of representations upon quantisation. This is also true about the (equivalent) discrete symmetries $(t,\f)\mapsto(-t,-\f)$ and $r_\pm\mapsto-r_\pm$. Both of these act on solution matrices as conjugation by $i\s^2$.

\subsection{\texorpdfstring{\SL2}{SL(2,R)} symmetries}
When we regard the solution as an \SL2 {\em matrix}, several natural discrete symmetries become visible. Given a solution, the inverse of the matrix gives a new solution. We can also consider the transpose of the matrix or multiply by $-\mathbb{I}$. Conjugation by Pauli matrices also produces more solutions. These transformations are discrete isometries of \SL2. Therefore, they will act as global symmetries on the space of classical {\em geodesic} solutions of the WZW model. Of these transformations, multiplication by $-\mathbb{I}$ and conjugation by $\s^3$ commute with the orbifolding. Hence, they will act on the space of states of the BTZ black hole as well.

\paragraph{Conjugation by Pauli Matrices.}
Conjugation by $i\s^2$ is part of the \SL2$\times$\SL2 symmetry of the WZW model and hence, does not lead to multiplicities of representations upon quantisation. Under its action on a solution, the diagonalised components of WZW currents evaluated along the solution transform as $J^{(2)}_\pm\mapsto-J^{(2)}_\pm$. For a BTZ geodesic having energy $E$ and angular momentum $L$, this implies $(E,L)\mapsto(-E,-L)$. On the other hand, conjugation by $\s^3$ is not part of the \SL2$\times$\SL2 symmetry -- it is an outer automorphism and leaves the diagonalised components $J^{(2)}_\pm$ invariant. Finally, conjugation by $\s^1$ is simply the conjugation by $i\s^2\s^3$.

\paragraph{Inverse and Transpose.}
The inverse $(g(\tau))^{-1}$ and transpose $(g(\tau))^t$ of a matrix $g(\tau)$ obtained by lifting a BTZ geodesic also project to BTZ geodesics. However, this is not true for generic $\s$-dependent solutions of the WZW model. This is easily seen by noting that $g\mapsto g^{-1}$ and $g\mapsto g^{t}$ do not preserve the WZW equations of motion $\del_\pm J_\mp=0$ (see Appendix~\ref{sec:Conventions} for conventions). However, combining the two operations preserves the equations of motion. In fact, it is equivalent to conjugation by $i\s^2$.

\paragraph{Multiplication by $-\mathbb{I}$.}
This transformation acts differently on lifts of different kinds of BTZ solutions. Consider a solution $g(\tau)$ of the WZW model obtained by lifting a timelike BTZ geodesic. Multiplying this by $-\mathbb{I}$ leaves it unchanged because such a solution is periodic, and its sign can be flipped by a translation $m\tau\mapsto m\tau+\pi$ of the affine parameter. However, for the {\em spacelike} geodesics, $g(\tau)$ and $-g(\tau)$ represent two distinct solutions of the \SL2 WZW model (which project to the same BTZ geodesic, albeit in different charts). We propose that this is the classical version of the statement that for each value $s^2+\frac{1}{4}$ of the quadratic Casimir, the corresponding principal continuous series representation (in the hyperbolic basis) has two states with identical energy and angular momentum quantum numbers~\cite{Mukunda}. Such a doubling is of course, absent for the discrete series.

\paragraph{Summary.}
We summarise the discussion by asking how many distinct geodesics exist for given values $(E,L,m^2)$ of energy, angular momentum and quadratic Casimir. For the sake of concreteness, we focus on the spacelike geodesic \eqref{eqn:SpacelikeExample} with $J^{(2)}_+>\frac{k|m|}{2}$ and $J^{(2)}_-<-\frac{k|m|}{2}$, which remains entirely in the $+D_1^+$ chart. Conjugation by $\s^3$ transforms this to a geodesic lying entirely in $+D_1^-$. Multiplying these two solutions by $-\mathbb{I}$ results in geodesics lying entirely in the charts $-D_1^+$ and  $-D_1^-$ respectively. Hence, upon quantisation, we expect 4 states corresponding to {\em spacelike} geodesics for given quantum numbers $(E,L,s^2+\frac{1}{4})$. On the other hand, because multiplication by $-\mathbb{I}$ leaves {\em timelike} geodesics invariant, we expect only 2 states corresponding to them for a given $(E,L,-j(j+1))$. The states of the string theory will be obtained by quantizing these geodesics and winding strings obtained by twisting them.

\section{Twisted sectors} \label{sec:TwistedSectors}
In this section, we shall study the twisted sector solutions following~\cite{NS,kraus,Esko}. The twisted sector solutions are defined by the equation
\be 
g_\Omega (\tau,\s+2\pi)= e^{\Omega \pi(r_+-r_-) \s^3}g_\Omega(\tau,\s)
e^{\Omega \pi(r_++r_-)\s^3}\,,\label{eqn:twistdefn}
\ee
with $\Omega$ defining the twisting parameter. It maybe observed -- in the parametrisation \eqref{eqn:param}, for example -- that the right hand side translates into the shift $\f\mapsto \f+2\pi\Omega$. Since $\f\sim\f+2\pi$ (due to the BTZ identification), the twisted solutions will be compatible with the orbifolding if $\Omega$ is an integer. More explicitly, given a solution $\tilde{g}$ of the WZW model such that $\tilde{g}(\tau,\s+2\pi)=\tilde{g}(\tau,\s)$, the following is a twisted sector solution:
\be
g_\Omega(\tau,\s)=e^{\frac{\Omega}{2}(r_+-r_-)\s^3 (\tau+\s)}\tilde g(\tau,\s) e^{-\frac{\Omega}{2} (r_++r_-)\s^3(\tau-\s)}\label{eqn:specflowdefn}
\ee
i.e., $g_\Omega$ satisfies the orbifolding condition \eqref{eqn:twistdefn}. 

The components of currents evaluated along the twisted solution \eqref{eqn:specflowdefn} are
\begin{subequations}\label{eqn:SpecFlowLabels}
\begin{align}
J^{(2)}_\pm&=\tilde J^{(2)}_\pm-\frac{k}{2}\Omega(r_+\mp r_-)\,,\label{eqn:SpecFlowLabels:Currents}\\
J^{(+)}_\pm&=e^{-\Omega(r_+\mp r_-)(\tau\pm\s)}\tilde J^{(+)}_\pm\,,\\
J^{(-)}_\pm&=e^{+\Omega(r_+\mp r_-)(\tau\pm\s)}\tilde J^{(-)}_\pm\,.
\end{align}
Here, $\tilde{J}_\pm$ denote currents evaluated on the solution $\tilde{g}$ before twisting. These relations imply that the Poisson brackets \eqref{eqn:HyperbolicBasis} are invariant under twisting. The stress tensor of the twisted solution evaluates to:
\be
\begin{aligned}
T_{\pm\pm}&=\tilde{T}_{\pm\pm}+\frac{k}{2}\Omega(\tilde{E}\pm\tilde{L})+\frac{k}{4}\Omega^2(r_+\mp r_-)^2\\
&=\tilde{T}_{\pm\pm}-\Omega(r_+\mp r_-)\tilde J^{(2)}_\pm+\frac{k}{4}\Omega^2(r_+\mp r_-)^2
\end{aligned}\label{eqn:SpecFlowLabels:StressTensor}
\ee
\end{subequations}
In the quantum theory, this implies transformations of the corresponding zero-mode eigenvalues
\be
L_0=\tilde{L}_0-\Omega(r_+\mp r_-)\tilde{J}^{(2)}_{\pm,0}+\frac{k}{4}\Omega^2(r_+\mp r_-)^2+N\,,\label{eqn:LowestWeightBTZ}
\ee
where the level $N$ is a non-negative integer. 
This differs from the $AdS_3$ case in an important way for the discrete series representations. Consider the spectrally flowed lowest weight discrete series $\hat{\mathcal{D}}_j^{+,w}$ in elliptic basis, for which
\be
L_0=\frac{-(j+\frac{1}{2})^2}{k-2}+\frac{1}{4(k-2)}+w\tilde{J}^{(0)}_{\pm,0}-\frac{k}{4}w^2+N\,,\qquad\tilde{J}^{(0)}_{\pm,0}\in\{j,j+1,j+2,\ldots\}\,.\label{eqn:LowestWeightAdS}
\ee
We have used $w$ to denote the $AdS_3$ spectral flow parameter. As is well known, this representation is identified with a highest weight discrete series having one less unit of spectral flow i.e.,
\be
\hat{\mathcal{D}}_j^{+,w}=\hat{\mathcal{D}}_{-\frac{k}{2}-j}^{-,w-1}\,.\label{eqn:DiscreteIdentify}
\ee
This may be verified by shifting $w$ in \eqref{eqn:LowestWeightAdS} and completing a square. However, this is not possible in \eqref{eqn:LowestWeightBTZ}. Hence, there is no such identification in the BTZ case.

The spacetime energy was defined in equation \eqref{eqn:Energy}; using the above, we see that the winding number $\Omega$ adds to the spacetime energy as
\be
E=\tilde{E}+\Omega M\,.\label{eqn:EnergyTwist}
\ee
On the other hand, for the spacetime angular momentum, it was pointed out by Hemming and Keski-Vakkuri~\cite{Esko} that for consistency with level matching on the worldsheet, we should use
\be
Q_\f=\int_0^{2\pi}\frac{d\s}{2\pi}\left(-\D_-J_+^{(2)} +\D_+ J_-^{(2)} + \frac{k}{2} J \del_\s \f \right)\label{eqn:TranslationOperatorTwist}
\ee 
to define the $\f$ translation operator. For the twisted solution, the angular momentum $kL=Q_\f$ evaluates to
\be
L=\tilde{L}-\frac{\Omega}{2}J\,.\label{eqn:AngularMometnumTwist} 
\ee
Note that only $t$ and $\f$ coordinates of the worldsheet change under twisting but the radial extent (in the BTZ geometry) of the worldsheet is not affected:
\be
t(\tau,\s)\mapsto t(\tau,\s)+\Omega\tau\,,\qquad\f(\tau,\s)\mapsto\f(\tau,\s)+\Omega\s\,,\qquad r^2(\tau)\mapsto r^2(\tau)\,.
\ee
This implies that if a geodesic $\tilde{g}(\tau)$ lies entirely in the region $r^2>0$ without CTCs, then the twisted (winding) string $g_\Omega$ also remains in the region without CTCs.

The definition of $\beta$ appearing in the geodesic equations \eqref{eqn:geodesic} may be extended to the twisted sector by replacing the $-\frac{1}{4}m^2$ appearing in it by $\frac{1}{2k}(T_{++}+T_{--})$. Rather remarkably, the condition  $\beta<0$ is now seen to be invariant under twisting: 
\be
\beta=ML^2+ELJ+(T_{++}+T_{--})\frac{J^2}{2k}=M\tilde L^2 + \tilde E\tilde L J+(\tilde T_{++}+\tilde T_{--})\frac{J^2}{2k}=\tilde{\beta}\,.
\ee
 
At this stage, we are left with a somewhat mixed situation. As far as the timelike geodesics are concerned, there is a single condition $\beta<0$, which eliminates the need to consider the regions with CTCs. It is also satisfactory that this condition is invariant under the spectral flow which produces the twisted sectors. However, there is an entire family of spacelike geodesics with $\beta>0$ which never penetrate the region with CTCs (those with $\alpha<0$ and upper sign choice in \eqref{eqn:spacelikeradial:positive}). Thus, it may be incorrect to impose $\beta<0$ as a condition on {\em all} geodesics to keep them out of the region with CTCs. Hence, a natural question is whether the Virasoro constraints remove geodesics that enter the region with CTCs.

\section{Virasoro conditions}
In this section, we will explore if a consistent excision of the states penetrating the unphysical region can occur by using the Virasoro conditions. Physical states in string theory, at the classical level, require the vanishing of the worldsheet stress tensor
\be
T_{++}+h_+ =0 = T_{--}+h_-\,,
\ee
where $h_{\pm}$ is the contribution from the `internal CFT'. We assume that the internal CFT is unitary and hence, $h_\pm\geq0$. Clearly, only timelike geodesics (and possibly null) with $T_{++}=-\frac{k}{4}m^2$ can be physical in the untwisted sector. However, the Virasoro constraints $-\frac{k}{4}m^2+h_\pm=0$ on timelike geodesics constraint only the quadratic Casimir $m^2$ but not their energy and angular momentum. Hence, they do not already prevent them from exploring the region with CTCs.

In the twisted sectors with twist $\Omega$, the Virasoro conditions become
\be
\tilde{T}_{\pm\pm}-\Omega(r_+\mp r_-)\tilde{J}^{(2)}_\pm+\frac{k}{4}\Omega^2(r_+\mp r_-)^2+h_\pm =0\,.\label{eqn:VirasoroConstraint}
\ee
By adding the Virasoro conditions, we get a relation for the energy
\be
k\Omega\tilde{E}=-2\tilde{T}_{\pm\pm}-\frac{kM\Omega^2}{2}-(h_++h_-)=k\Omega(E-\Omega M)\,.
\ee
We see that for fixed $h_\pm$ and $\Omega>0$, the energies of spectral flows of timelike geodesics ($\tilde{T}_{\pm\pm}<0$) are bounded below
\be
E>\frac{M\Omega}{2}-\frac{h_++h_-}{k\Omega}
\ee
as in the case of $AdS_3$ studied in~\cite{mog}. Clearly, states of the continuous series ($\tilde{T}_{\pm\pm}>0$) will step in to obviate this restriction on energy. The difference with the $AdS_3$ case is the presence of the black hole mass factor. As in $AdS_3$, the states of the continuous series (spacelike geodesics) scatter out to the boundary. Thus, these can play the role of operators of the dual CFT~\cite{Ooguri}.

The solution obtained by $\Omega$ units of spectral flow on a timelike geodesic does not enter the region with CTCs, if it satisfies $\beta<0$. Rewriting this condition using the Virasoro constraints \eqref{eqn:VirasoroConstraint}, we find
\be
\left(h_+-h_--\frac{k}{2}J\Omega^2\right)\left[(h_+-h_-)M+J(h_++h_-)\right]<\frac{k}{2}m^2J(h_+-h_-)\,.\label{eqn:BetaVirasoro}
\ee
Clearly, this condition is not automatic for a generic internal CFT. In fact, we find that physical states of the string theory which satisfy\footnote{This was derived assuming $(h_+-h_-)\Omega>0$.}
\be
E < \frac{M}{k\Omega}(h_+ - h_-) + \frac{J}{k\Omega}(h_+ + h_-) - \frac{J\Omega M}{2}-\frac{J^2\Omega}{2} \left(\frac{h_+ + h_-}{h_+ - h_-} \right) - \frac{h_+ + h_-}{k \Omega} + \frac{M\Omega}{2}\,,
\ee 
do pass into the region with CTCs. However, it may be possible to choose an internal CFT so that $\beta<0$ is automatic. For example, $h_+=h_-$ automates $\beta<0$ but this is perhaps, too strong a requirement. In any case, \eqref{eqn:BetaVirasoro} does not prevent physical timelike geodesics from entering the region with CTCs. It only ensures that physical {\em winding} strings do not explore this region. Since $\beta$ is invariant under twisting, it also prevents geodesics that lead to physical winding strings from exploring these regions.

All these considerations do not apply to spacelike geodesics since there are spacelike geodesics that exist entirely in the region with CTCs. Thus, in conclusion, the Virasoro conditions do not help truncate the space of states to those that remain in a causally sensible region of the BTZ geometry.

\section{Torus partition function and modular invariance}
In this section, we show that the modular invariant partition function for global $AdS_3$ constructed in~\cite{Israel} contains all the states of the BTZ black hole that were discussed previously.\footnote{A similar remark has already been made by Eberhardt~\cite{Eberhardt} in the tensionless limit.} The fact that these are two different spacetimes arises when we try to identify the spacetime quantum numbers from the partition function.

We start with the partition function for $AdS_3$ as written in~\cite{Israel} (see Appendix~\ref{sec:Notation} for notation):
\be
\begin{aligned}
Z&=4\sqrt{\tau_2}(k-2)^{\frac32}\int_0^1d^2s\;\int_0^1d^2t\;\frac{e^{\frac{2\pi}{\tau_2}({\text{Im}}(s_1\tau-s_2))^2}}{|\vartheta_1(s_1\tau-s_2| \tau)|^2}\\
&\quad\times\sum_{m,w,m',w'\in\mathbb{Z}} \z\oao{w+s_1-t_1}{m+s_2-t_2}(k)\;\z\oao{w'+t_1}{m'+t_2}(-k)
\end{aligned}\label{eqn:ppsedbtzsl2}
\ee
In what follows, we first expand this partition function as a power series in $q$ to identify the states. This allows us to identify the spacetime isometry currents for these states and make contact with the analysis of~\cite{NS}. Finally, we compare this identification of the currents with the corresponding relation for $AdS_3.$

\subsection{Series expansion of the partition function}
We follow the procedure detailed in~\cite{Israel} to expand the partition function \eqref{eqn:ppsedbtzsl2}. First, we expand the factor $|\vartheta_1|^{-2}$ as explained in Appendix~\ref{sec:Notation} and trade off $(m,m')$ for $(n,n')$ by Poisson resummation. The exponent in $s_1$ can be linearised by introducing a Gaussian integral over $s$  allowing us to integrate $s_1$ out and obtain:
\be 
\begin{aligned}
Z&=\frac{4(k-2)}{\pi}\sum_{\substack{w,w',n,n'\in\mathbb{Z}\\q,\bar{q},N,\bar{N}}}\int_0^1dt_2\int_0^1ds_2\;\exp\left[2\pi it_2(n-n')+2\pi is_2(\bar{q}-q-n)\right]\\
&\quad\times\int_0^1dt_1\int_{-\infty}^{\infty}ds\left[\frac{1}{2is+q+\bar{q}+1+k(w-t_1)}-\frac{e^{-2\pi\tau_2(2is+q+\bar{q}+1+k(w-t_1))}}{2is+q+\bar{q}+1+k(w-t_1)}\right]\\
&\quad\times\exp\left[2\pi i\tau_1(n(w-t_1)+n'(w'+t_1)+N-\bar{N})\right]\\
&\quad\times\exp\left[-2\pi\tau_2\left(\frac{2s^2}{k-2}-\frac{1}{4}+\frac{n^2}{2k}-\frac{n'^2}{2k}+\frac{k(w-t_1)^2}{2}-\frac{k(w'+t_1)^2}{2}+N+\bar{N}\right)\right]
\end{aligned} \label{eqn:sl2expansion}
\ee
The integrals over $t_2$ and $s_2$ lead to the constraints $n=n'=\bar{q}-q$. Of the two integrals in the second line, we change variables in the second as
\be
\tilde{s}=s+\frac{i}{2}(k-2)\,,\qquad\quad\tilde{w}=w+1\,,\qquad\quad\tilde{N}=N+q\,,\qquad\quad\tilde{\bar{N}}=\bar{N}+\bar{q}\,.\label{eqn:SpecFlowPartFunc}
\ee
This change of variables corresponds to a spectral flow if \eqref{eqn:ppsedbtzsl2} is interpreted as the global $AdS_3$ partition function. Hence, the sum over $(q,\bar{q},N,\bar{N})$ differs from that over $(q,\bar{q},\tilde{N},\tilde{\bar{N}})$ only in the signs of zero mode contributions~\cite{mog2} (denoted $P_{\pm,0}$ in \eqref{eqn:degeneracymatch:qbarq}) to $(q,\bar{q})$. The factor $e^{-2\pi\tau_2(2is+q+\bar{q}+1+k(w-t_1))}$ is now completely absorbed into the new variables. However, the $\tilde{s}$ integral now runs along $\mathbb{R}+\frac{i}{2}(k-2)$. This differs from the integral in the first term by residues of poles in the strip enclosed by $\mathbb{R}+\frac{i}{2}(k-2)$ and $\mathbb{R}$. Collecting these residues, we obtain the net result:
\be 
\begin{aligned}
Z&=8i(k-2)\sum_{\substack{w,w',n,n'\in\mathbb{Z}\\N,\bar{N}}}\delta_{n,n'}\int_0^1dt_1\Biggl\{\int_0^{\infty}ds\;\rho(s)\exp\left[-2\pi\tau_2\left(\frac{2s^2+1/2}{k-2}\right)\right]\\
&\quad+\sum_{q,\bar{q}}\delta_{n,\bar{q}-q}\exp\left[-2\pi\tau_2\left(\frac{-2j(j+1)}{k-2}\right)\right]_{\frac{1}{2}<-j=\frac{q+\bar{q}}{2}+\frac{k}{2}(w-t_1)<\frac{k-1}{2}}\Biggr\}\\
&\quad\times\exp\left[2\pi i\tau_1(n(w-t_1)+n'(w'+t_1)+N-\bar{N})\right]\\
&\quad\times\exp\left[-2\pi\tau_2\left(\frac{n^2}{2k}-\frac{n'^2}{2k}+\frac{k(w-t_1)^2}{2}-\frac{k(w'+t_1)^2}{2}-\frac{3k}{12(k-2)}+N+\bar{N}\right)\right]\label{eqn:PartFuncExpand}
\end{aligned}
\ee
Here, the second line corresponds to states of the discrete series. The first line is obtained by combining the $s$ and $\tilde{s}$ integrals over $\mathbb{R}$ and {\em further} combining the contributions from positive and negative values of the integration variable. This denotes contribution from the continuous series representations, whose density of states for a given quadratic Casimir $s^2+\frac{1}{4}$ is
\be
\rho(s)=\frac{1}{\p i}{\text{Re}}\left[\sideset{}{^+}\sum_{q,\bar{q}}\frac{\delta_{n,\bar{q}-q}}{2is+q+\bar{q}+1+k(w-t_1)}-\sideset{}{^-}\sum_{q,\bar{q}}\frac{\delta_{n,\bar{q}-q}}{2is+q+\bar{q}-1+k(w-t_1)}\right]\,,
\ee
where the superscripts on sums indicate signs of zero mode contributions to $(q,\bar{q})$. They lead to divergences that need to be regularised~\cite{mog2} to obtain
\begin{multline}
\sum_{N,\bar{N}}\rho(s)=\sum_{\substack{P_{\pm,p},P^+_{\pm,p},P^-_{\pm,p}=0\\p\in\{0,1,2,3,\ldots\}}}^\infty\Biggl\{\frac{\ln\e}{\p i}+\frac{1}{4\p i}\frac{d}{ids}\ln\Biggl[\frac{\G(\frac{1}{2}-is+\frac{|n+q'-\bar{q}'|+q'+\bar{q}'}{2}+\frac{k}{2}(w-t_1))}{\G(\frac{1}{2}+is+\frac{|n+q'-\bar{q}'|+q'+\bar{q}'}{2}+\frac{k}{2}(w-t_1))}\\
\times\frac{\G(\frac{1}{2}-is+\frac{|n+q'-\bar{q}'|-q'-\bar{q}'}{2}-\frac{k}{2}(w-t_1))}{\G(\frac{1}{2}+is+\frac{|n+q'-\bar{q}'|-q'-\bar{q}'}{2}-\frac{k}{2}(w-t_1))}\Biggr]+\mathcal{O}(\e)\Biggr\}\,,\label{eqn:DensityOfStates}
\end{multline}
where $(q',\bar{q}')$ denote the expressions \eqref{eqn:degeneracymatch:qbarq} for $(q,\bar{q})$ void of the zero mode contributions (i.e., $P_{+,0}=0=P_{-,0}$).

We can now identify the $J^{(2)}_{\pm,0}$ eigenvalues $\tilde{J}_\pm$ (and $J_\pm$) before (and after) spectral flow respectively as
\be
\begin{pmatrix}
\tilde{J}_+\\J_+\\\tilde{J}_-\\J_-
\end{pmatrix}=\frac{k}{2}\begin{pmatrix}
-\cosh\q_+&\sinh\q_+&0&0\\
-\sinh\q_+&\cosh\q_+&0&0\\
0&0&-\cosh\q_-&\sinh\q_-\\
0&0&-\sinh\q_-&\cosh\q_-
\end{pmatrix}\begin{pmatrix}
w'+t_1-\frac{n'}{k}\\
w-t_1+\frac{n}{k}\\
w'+t_1+\frac{n'}{k}\\
w-t_1-\frac{n}{k}
\end{pmatrix},\label{eqn:CurrentsIdentify}
\ee
where $e^{-\q_\pm}=\D_\mp=r_+\mp r_-$. This identification ensures that $L_0\pm\bar{L}_0$ for primaries (i.e., $N=0=\bar{N}$) read off from \eqref{eqn:PartFuncExpand} have exactly the forms
\be
\begin{split}
\begin{aligned}
L_0-\bar{L}_0&=\frac{1}{k}(-\tilde{J}_+^2+J_+^2)-\frac{1}{k}(-\tilde{J}_-^2+J_-^2)\,,\\
L_0+\bar{L}_0&=\frac{1}{k}(-\tilde{J}_+^2+J_+^2)+\frac{1}{k}(-\tilde{J}_-^2+J_-^2)+\frac{2(s^2+\frac{1}{4})}{k-2}\,,
\end{aligned}    
\end{split}\qquad\quad
\begin{split}
J_\pm&=\tilde{J}_\pm-\frac{k}{2}\Omega\D_\mp\,,
\end{split}\label{eqn:NatSatProposal}
\ee
which are as expected from the classical result \eqref{eqn:SpecFlowLabels:StressTensor} for a spectrally flowed primary. The same expressions for $L_0\pm\bar{L}_0$ were also obtained in the quantised CFT by Natsuume and Satoh~\cite{NS}. Along with the expected form of $L_0\pm\bar{L}_0$, this identification also ensures the following:
\begin{itemize}
\item The spectral flow parameter turns out to be $\Omega=-(w+w')$ as is clear from
\be
J_+-\tilde{J}_+=\frac{k}{2}\left(w+w'+\frac{n-n'}{k}\right)\D_-\,,\quad J_--\tilde{J}_-=\frac{k}{2}\left(w+w'-\frac{n-n'}{k}\right)\D_+\,,
\ee
using the constraint $n=n'$ appearing in \eqref{eqn:PartFuncExpand}. This constraint ensures that the winding number along the non-compact $t$ direction of the BTZ geometry is zero.
\item While analysing geodesics on the BTZ geometry, we expressed the energy and angular momentum of a geodesic as linear combinations \eqref{eqn:Energy} of WZW currents. Evaluating these using the identification \eqref{eqn:CurrentsIdentify}, we find that
\begin{subequations}
\begin{align}
k\tilde{L}&=-\D_-\tilde{J}_++\D_+\tilde{J}_-=-n-\frac{k}{2}(w+w')J\,,\\
k\tilde{E}&=-\D_-\tilde{J}_+-\D_+\tilde{J}_-=-\frac{k}{2}(w-w'-2t_1)+\frac{k}{2}(w+w')M\,, \label{eqn:PreTwistIdentifyEnergy}
\end{align}\label{eqn:PreTwistIdentify}
\end{subequations} 
where $J=2r_+r_-$ and $M=r_+^2+r_-^2$ are the black hole angular momentum and mass, respectively. Setting $\Omega=-(w+w')=0$ we observe that angular momentum is quantised, while the energy is an arbitrary real number. This is exactly as expected for geodesics on the BTZ geometry because the $\phi$ coordinate is $2\p$-periodic while the $t$ coordinate is non-compact. So, we may identify the states appearing in \eqref{eqn:PartFuncExpand} with $w+w'=0$ as those obtained by quantizing the geodesics of the BTZ black hole.
\item As discussed previously, the energy and angular momentum of a solution twisted by $\Omega$ units are related to those before twisting as in \eqref{eqn:EnergyTwist} and \eqref{eqn:AngularMometnumTwist}. The identification \eqref{eqn:CurrentsIdentify} also reproduces these results
\begin{subequations}
\begin{align}
kL&=-\D_-J_++\D_+J_-+\frac{k}{2}\Omega J=-n=k\left(\tilde{L}-\frac{\Omega}{2}J\right)\,,\\
kE&=-\D_-J_+-\D_+J_-=-\frac{k}{2}(w-w'-2t_1)-\frac{k}{2}(w+w')M=k(\tilde{E}+\Omega M)\,.
\end{align}\label{eqn:PostTwsistIdentify}
\end{subequations}
Notice that it is the generator $Q_\f$ defined in \eqref{eqn:TranslationOperatorTwist} that is quantised rather than the combination $-\D_-J_++\D_+J_-$. Hence, our identification \eqref{eqn:CurrentsIdentify} is consistent with the Noether ambiguity pointed out by Hemming and Keski-Vakkuri~\cite{Esko}.
\item From our analysis of discrete symmetries in Section~\ref{sec:DiscreteSymmetries}, we concluded a doubling of continuous series representations relative to the discrete series. However, considering winding strings {\em and} descendants as in Section~\ref{sec:TwistedSectors}, an identification similar to \eqref{eqn:DiscreteIdentify} was found to be missing in the BTZ case. This revealed a doubling of discrete series representations in BTZ relative to $AdS_3$. Hence, it is satisfactory that our BTZ interpretation of the $AdS_3$ partition function follows without changing the relative degeneracy between discrete and continuous series contributions in \eqref{eqn:PartFuncExpand}.
\end{itemize}
Having identified the physical quantum numbers $(E, L)$, we can rewrite the density of states \eqref{eqn:DensityOfStates} in terms of them as
\begin{multline}
\sum_{N,\bar{N}}\rho(s)=\sum_{\substack{P_{\pm,p},P^+_{\pm,p},P^-_{\pm,p}=0\\p\in\{0,1,2,3,\ldots\}}}^\infty\frac{1}{4\p i}\frac{d}{ids}\ln\Biggl[\frac{\G(\frac{1}{2}-is+\frac{|kL-q'+\bar{q}'|+q'+\bar{q}'}{2}-\frac{k}{2}(E-\frac{\Omega}{2}(M-1)))}{\G(\frac{1}{2}+is+\frac{|kL-q'+\bar{q}'|+q'+\bar{q}'}{2}-\frac{k}{2}(E-\frac{\Omega}{2}(M-1)))}\\
\times\frac{\G(\frac{1}{2}-is+\frac{|kL-q'+\bar{q}'|-q'-\bar{q}'}{2}+\frac{k}{2}(E-\frac{\Omega}{2}(M-1)))}{\G(\frac{1}{2}+is+\frac{|kL-q'+\bar{q}'|-q'-\bar{q}'}{2}+\frac{k}{2}(E-\frac{\Omega}{2}(M-1)))}\Biggr]\,,
\end{multline}
where we have set $\e=0$ after omitting the $\ln\e$ divergence for the sake of brevity. It is interesting that this expression does not involve the black hole angular momentum $J$.

To summarise, we have shown that the primaries (states with $N=0=\bar{N}$) read off from the partition function \eqref{eqn:ppsedbtzsl2} are in one-to-one correspondence with quantised geodesics on the BTZ black hole and string worldsheets winding around its $\phi$ direction, obtained by twisting the geodesics. 

Now, let us consider the states with non-zero $(N,\bar{N})$. As described in Appendix~\ref{sec:Notation}, the quantum numbers $(N,\bar{N})$ and $(q,\bar{q})$ arise from expansion of the factor $|\vartheta_1|^{-2}$ present in \eqref{eqn:ppsedbtzsl2}. This factor includes six terms -- three holomorphic and three anti-holomorphic -- along with two factors that contribute only to $(q,\bar{q})$ but not $(N,\bar{N})$. In what follows, we attempt to understand these quantum numbers using the vertex operator constructions proposed in~\cite{Satoh,Hemming}.

\subsection{Vertex operators}
We begin this section with a brief review of the free field realisation of $\hat{\mathfrak{sl}}(2)_k$ introduced by Satoh~\cite{Satoh}. Then, we describe the vertex operators written by Satoh~\cite{Satoh} and Hemming~\cite{Hemming}, corresponding to the primary states read off from the partition function. We shall use these vertex operators to propose an interpretation for the quantum numbers $(q,\bar{q},N,\bar{N})$ and therefore the factor $|\vartheta_1|^{-2}$ appearing in \eqref{eqn:ppsedbtzsl2}.

The free field realisation of the $\hat{\mathfrak{sl}}(2)_k$ algebra provided by Satoh~\cite{Satoh} consists of three chiral bosonic fields $X_a$ (with $a\in\{0,1,2\}$) satisfying OPEs
\be
X_a(z)X_b(w)\sim-\eta_{ab}\ln(z-w)\,,\qquad\qquad\eta_{ab}={\text{diag}}(-1,1,1)\,.
\ee
The holomorphic $\hat{\mathfrak{sl}}(2)_k$ algebra in hyperbolic basis (see Appendix~\ref{sec:Conventions}) and its Sugawara stress tensor is realised in terms of these fields as
\begin{subequations}
\begin{align}
iJ_+^{(\pm)}&=e^{\mp\sqrt{\frac{2}{k}}(X_0-X_1)}\left(\sqrt{\frac{k}{2}}\partial X_0\mp\sqrt{\frac{k-2}{2}}\partial X_2\right)\,,\\
iJ_+^{(2)}&=\sqrt{\frac{k}{2}}\partial X_1\,,\\
T_{++}&=-\frac{1}{2}\eta^{ab}\partial X_a\partial X_b-\frac{1}{\sqrt{2(k-2)}}\partial^2 X_2\,.
\end{align}
\end{subequations}
Using this free field realisation, we may verify that the holomorphic vertex operator
\be
V^C_{j,\tilde{J}_+,J_+}=\exp\left[\left(i\tilde{J}_+X_0-iJ_+X_1\right)\sqrt{\frac{2}{k}}+jX_2\sqrt{\frac{2}{k-2}}\right]\label{eqn:CtsVertexOperators}
\ee
has $L_0$ eigenvalue $\frac{-j(j+1)}{k-2}+\frac{1}{k}(-\tilde{J}_+^2+J_+^2)$ and $J^{(2)}_{+,0}$ eigenvalue $J_+$. Thus, the vertex operator $V^C_{-\frac{1}{2}+is,\tilde{J}_+,\tilde{J}_+-\frac{k}{2}\Omega\D_-}$ corresponds to the holomorphic part of a state obtained by a twist of $\Omega$ units from a continuous series state with $J^{(2)}_{\pm,0}$ eigenvalues $\tilde{J}_\pm$.

On the other hand, the interchange $X_0\leftrightarrow iX_2$ yields another free field realisation of the same $\hat{\mathfrak{sl}}(2)_k$ algebra. It was pointed out by Hemming~\cite{Hemming} that this realisation can be used to write the vertex operators $V^D_{j,\tilde{J}_+,J_+}$ corresponding to holomorphic parts of discrete series states. These are obtained from \eqref{eqn:CtsVertexOperators} by the same interchange $X_0\leftrightarrow iX_2$.

Similarly, we may use anti-holomorphic parts $\tilde{X}_a$ (with $a\in\{0,1,2\}$) of the bosonic fields to obtain a free field realisation of the anti-holomorphic current algebra and corresponding vertex operators. The vertex operators corresponding to states read off from the partition function \eqref{eqn:ppsedbtzsl2} are then $V_{j,\tilde{J}_+,J_+}\tilde{V}_{j,\tilde{J}_-,J_-}$ with
\be
\tilde{J}_\pm=\frac{\frac{k}{2}(w-w^\prime-2t_1)\pm n}{2\D_\mp}-\frac{k}{4}(w+w^\prime)\D_\mp\,,\qquad\quad J_\pm=\tilde{J}_\pm+\frac{k}{2}(w+w^\prime)\D_\mp\,.\label{eqn:CurrentsFromPartFunc}
\ee
We shall use this to propose an interpretation of the denominator $|\vartheta_1|^2$ appearing in \eqref{eqn:ppsedbtzsl2}.

To begin with, the factors in $|\vartheta_1|^{-2}$ do not represent the modes $(J^{(2)}_{\pm,n\leq0},J^{(+)}_{\pm,n<0},J^{(-)}_{\pm,n<0})$. If this were the case, then the hyperbolic basis commutation relations \eqref{eqn:HyperbolicBasis:ifactor} would imply that they contribute $(iq, i\bar{q})$ additively to $(J^{(2)}_{+,0},J^{(2)}_{-,0})$ eigenvalues, without any $(\D_-,\D_+)$ factors. Instead they contribute $\pm\frac{\bar{q}-q}{2\D_\mp}$ to $J^{(2)}_{\pm,0}$ eigenvalues as is clear from \eqref{eqn:CurrentsFromPartFunc} and the constraint $n=\bar{q}-q$ from \eqref{eqn:PartFuncExpand}. Furthermore, the second line of \eqref{eqn:PartFuncExpand} makes it clear that the factor $|\vartheta_1|^{-2}$ also contributes $-\frac{q+\bar{q}}{2}$ to $j$-values of the discrete series. Hence, the denominator $|\vartheta_1|^2$ has to be interpreted differently for states of the discrete and continuous series.

An interpretation of the factors in $|\vartheta_1|^{-2}$ that is consistent with these observations is as follows. In terms corresponding to the continuous series in \eqref{eqn:PartFuncExpand}, the six factors in the infinite product from $|\vartheta_1|^{-2}$ can be attributed to the operators
\begin{subequations}
\begin{align}
J^{(2)}_{+,-p}\,,&& A^\pm_{+,-p}&=J^{(2)}_{+,-p}V^C_{0,\frac{\mp1}{2\D_-},\frac{\mp1}{2\D_-}}\tilde{V}^C_{0,\frac{\pm1}{2\D_+},\frac{\pm1}{2\D_+}}\,,\\
J^{(2)}_{-,-p}\,,&& A^\pm_{-,-p}&=J^{(2)}_{-,-p}V^C_{0,\frac{\pm1}{2\D_-},\frac{\pm1}{2\D_-}}\tilde{V}^C_{0,\frac{\mp1}{2\D_+},\frac{\mp1}{2\D_+}}\,,
\end{align}
\end{subequations}
for each $p\in\{1,2,3,...\}$. These operators contribute to both $(q,\bar{q})$ and $(N,\bar{N})$. On the other hand, the two factors in $|\vartheta_1|^{-2}$ that diverge as $s_1,s_2\to0$ are due to
\be
A^+_{+,0}=V^C_{0,\frac{-1}{2\D_-},\frac{-1}{2\D_-}}\tilde{V}^C_{0,\frac{1}{2\D_+},\frac{1}{2\D_+}}\,,\qquad\qquad\qquad A^+_{-,0}= V^C_{0,\frac{1}{2\D_-},\frac{1}{2\D_-}}\tilde{V}^C_{0,\frac{-1}{2\D_+},\frac{-1}{2\D_+}}\,.
\ee
These operators contribute only to $J^{(2)}_{\pm,0}$ (via $q$ and $\bar{q}$) and not to $L_0\pm\bar{L}_0$, just like the corresponding factors in $|\vartheta_1|^{-2}$. This interpretation reproduces the exact degeneracies of $(q,\bar{q})$ and $(N,\bar{N})$ as they appear in \eqref{eqn:PartFuncExpand}. To see this, note that the contribution of these operators to $\tilde{J}_\pm$ and $J_\pm$ is precisely $\mp\frac{q-\bar{q}}{2\D_\mp}$ where
\begin{subequations}\label{eqn:qbarq}
\begin{align}
q&=\#A^+_{+,0}+\sum_{p=1}^\infty(\#A^+_{+,-p}-\#A^-_{+,-p})\,,\\
\bar{q}&=\#A^+_{-,0}+\sum_{p=1}^\infty(\#A^+_{-,-p}-\#A^-_{-,-p})\,,
\end{align}
\end{subequations}
and their contribution to $(N,\bar{N})$ is also as expected from the expansion of $|\vartheta_1|^{-2}$ (see Appendix~\ref{sec:Notation}) viz.,
\begin{subequations}\label{eqn:NbarN}
\begin{align}
N&=\sum_{p=1}^\infty p(\#J^{(2)}_{+,-p}+\#A^+_{+,-p}+\#A^-_{+,-p})\,,\\
\bar{N}&=\sum_{p=1}^\infty p(\#J^{(2)}_{-,-p}+\#A^+_{-,-p}+\#A^-_{-,-p})\,.
\end{align}
\end{subequations}
Comparing expressions \eqref{eqn:qbarq} and \eqref{eqn:NbarN} with \eqref{eqn:degeneracymatch} confirms that our interpretation reproduces degeneracies appearing in the partition function. 

As already explained, the same interpretation does not apply to $|\vartheta_1|^{-2}$ multiplying terms corresponding to the discrete series in \eqref{eqn:PartFuncExpand}. The various factors in $|\vartheta_1|^{-2}$ are now to be interpreted as arising due to the following operators
\begin{subequations}
\begin{align}
J^{(2)}_{+,-p}\,,&& A^\pm_{+,-p}&=J^{(2)}_{+,-p}V^D_{\frac{\mp1}{2},\frac{\mp1}{2\D_-},\frac{\mp1}{2\D_-}}\tilde{V}^D_{\frac{\mp1}{2},\frac{\pm1}{2\D_+},\frac{\pm1}{2\D_+}}\,,\\
J^{(2)}_{-,-p}\,,&& A^\pm_{-,-p}&=J^{(2)}_{-,-p}V^D_{\frac{\mp1}{2},\frac{\pm1}{2\D_-},\frac{\pm1}{2\D_-}}\tilde{V}^D_{\frac{\mp1}{2},\frac{\mp1}{2\D_+},\frac{\mp1}{2\D_+}}\,,\\
&&A^+_{\pm,0}&=V^D_{\frac{-1}{2},\frac{\mp1}{2\D_-},\frac{\mp1}{2\D_-}}\tilde{V}^D_{\frac{-1}{2},\frac{\pm1}{2\D_+},\frac{\pm1}{2\D_+}}\,,
\end{align}
\end{subequations}
for each $p\in\{1,2,3,...\}$. This interpretation correctly ensures the appearance of $\frac{q+\bar{q}}{2}$ in the discrete series $j$-values.

In summary, we have shown that the $AdS_3$ partition function \eqref{eqn:ppsedbtzsl2} contains quantised geodesics of the BTZ black hole and states obtained by twisting them. The same partition function also captures the doubling of the continuous series and the non-degeneracy of the discrete series. In addition, we have also shown that the spectrum of primaries agree with vertex operator constructions. All in all, this makes a strong case that the $AdS_3$ states carry over into the BTZ spectrum.

\subsection{Comparison with \texorpdfstring{$AdS_3$}{AdS\_3}}
The partition function \eqref{eqn:ppsedbtzsl2} is exactly the one written by Israel et al.~\cite{Israel} for $AdS_3$. So, let us examine the difference in interpretation that also revels the BTZ spectrum it encodes.

The conformal weights for global $AdS_3$, where the elliptic timelike component $J^{(0)}_{\pm,0}$ (see Appendix~\ref{sec:Conventions}) is diagonalised, can be written as
\be
\begin{split}
\begin{aligned}
L_0-\bar{L}_0&=\frac{1}{k}(\tilde{J}_+^2-J_+^2)-\frac{1}{k}(\tilde{J}_-^2-J_-^2)\,,\\
L_0+\bar{L}_0&=\frac{1}{k}(\tilde{J}_+^2-J_+^2)+\frac{1}{k}(\tilde{J}_-^2-J_-^2)+\frac{2(s^2+\frac{1}{4})}{k-2}\,,
\end{aligned}    
\end{split}\qquad\quad
\begin{split}
J_\pm&=\tilde{J}_\pm\pm\frac{k}{2}\Omega\,,
\end{split}
\ee
where $\tilde{J}_\pm$ (and $J_\pm$) now denote $J^{(0)}_{\pm,0}$ eigenvalues before (and after) spectral flow by $\Omega$ units. Comparing this with the coefficients of $2\p i\tau_1$ and $-2\p\tau_2$ in the exponent of \eqref{eqn:PartFuncExpand} we find:
\begin{subequations}\label{eqn:CurrentsIdentifyAdS3}
\begin{align}
\tilde{J}_+&=\frac{k}{2}\left(w-t_1+\frac{n}{k}\right)\,,\\
J_+&=-\frac{k}{2}\left(w'+t_1-\frac{n'}{k}\right)\,,\\
\tilde{J}_-&=-\frac{k}{2}\left(w-t_1-\frac{n}{k}\right)\,,\\
J_-&=\frac{k}{2}\left(w'+t_1+\frac{n'}{k}\right)\,.
\end{align}
\end{subequations}
This is the correct identification because it reproduces the correct quantisation of spacetime currents. The spectral flow parameter as calculated from both $J_+-\tilde{J}_+$ and $J_--\tilde{J}_-$ match and turn out to be $\Omega=-(w+w').$ This follows after using the constraint $n=n'$, which implies that there is no winding along the time direction of $AdS_3$ in global coordinates. Now, we may verify that for states with $w+w'=0$, the $AdS_3$ energy $\tilde{J}_+-\tilde{J}_-$ is an arbitrary real number while the angular momentum $\tilde{J}_++\tilde{J}_-$ is quantised.

It is important that the direction conjugate to the current $(J^{(0)}_+,J^{(0)}_-)$ in the $AdS_3$ geometry is not the direction conjugate to $(J^{(2)}_+,J^{(2)}_-)$, along which the BTZ orbifolding has been performed. We claim that this is reflected in the fact that there is no choice of $\q_+,\q_-\in\mathbb{C}$ that can be used in \eqref{eqn:CurrentsIdentify}, to get the $AdS_3$ currents \eqref{eqn:CurrentsIdentifyAdS3}. Instead, the matrix in \eqref{eqn:CurrentsIdentify} needs to be replaced by a rotation matrix as in \eqref{eqn:CurrentsIdentifyAdS3}, to obtain the $AdS_3$ currents.

\section{Summary and Discussion}
In this work, we revisited the spectrum of the BTZ black hole CFT first discussed in~\cite{NS}, starting with a careful study of the geodesics. The space of classical solutions, appropriately quantised, will provide a basis of states to construct the current algebra representations. 

We first identified conditions on solutions that restrict them to regions of the BTZ geometry without CTCs. Such a condition should be necessary because in regions with CTCs, the time coordinate ceases to be single-valued on the worldsheet. This will prevent a lightcone gauge choice and hence lead to difficulties in constructing a well-defined spacetime spectrum.

The continuous series states include worldsheets that reach the boundary of $AdS_3$ and thus represent operators of the dual CFT. It will be interesting to explore these operators in a putative dual such as the symmetric product CFT. The classical analysis revealed natural time scales associated with these geodesics, which could be expressed in terms of the currents. Therefore, we can expect that these time scales appear in the correlators of the corresponding operators in the dual CFT. This may lead to insights about the black hole interior~\cite{vijay,HongLiu}.

Lifting the geodesics to classical solutions of the \SL2 WZW model revealed more discrete symmetries of the space of geodesics than the ones obvious from the BTZ point of view. The study of discrete symmetries was used to predict multiplicities of quantum numbers in the quantised sigma model on the BTZ background. Among the discrete symmetries was multiplication by $-\mathbb{I}$. It revealed a classical version of the doubling~\cite{Mukunda} of continuous series representations of \SL2 in the hyperbolic basis. A similar doubling was also found in the $2$-dimensional Lorentzian black hole~\cite{kpy}. In that context, however, it was time-reversal symmetry that motivated this doubling. Another important discrete symmetry was $(r^2-r_+^2,t)\leftrightarrow(r_-^2-r^2,\f)$. As proposed by Natsuume and Satoh~\cite{NS}, orbifolding with respect to this T-duality could be interpreted as truncating the BTZ geometry to exclude the region with closed timelike curves. 

We also studied solutions of the sigma model obtained by twisting the geodesics. We noted that solutions obtained by twisting geodesics that do not explore the region with CTCs remain in the causally sound region. Also, the $\beta<0$ condition was invariant under twisting once the Noether ambiguity was taken into account. However, $\beta<0$ is too strong a constraint to impose on all solutions because many winding strings obtained by twisting spacelike geodesics do not satisfy $\beta<0$; yet, they stay out of the region with CTCs and may also satisfy Virasoro conditions.

While studying twisted sectors, we also noted that the highest weight discrete series does not map to another lowest weight discrete series under spectral flow in the hyperbolic basis. This is unlike the case in $AdS_3$ and implies a doubling of the discrete series, just like the continuous series. This fits nicely with the fact that BTZ states may be read off from the $AdS_3$ partition function without changing the relative contributions to the partition function from continuous and discrete series representations.

The quantum spectrum of the BTZ geometry is expected to contain states obtained by quantising the geodesics. These were already identified through a coset construction by~\cite{NS}. We then showed that the partition function of the $AdS_3$ sigma model as constructed in~\cite{Israel} contains all the states identified classically. Expanding this partition function in a $q$-series, we showed that it is possible to reinterpret the quantum numbers of primaries read off from this expansion to match those calculated by Natsuume and Satoh~\cite{NS}. This interpretation reproduced correct geometric features of the spectrum, such as angular momentum quantisation and the absence of winding around the non-compact direction. We also verified its consistency with the Noether ambiguity~\cite{Esko} encountered in the study of twisted sectors. The descendants of these primaries and their degeneracies could also be understood using free field realisations of the $\hat{\mathfrak{sl}}(2)_k$ algebra introduced by Satoh and Hemming~\cite{Satoh,Hemming}. 

The fact that the same partition function could describe two distinct spacetimes deserves some comment. The BTZ interpretation required the introduction of spacetime moduli $r_\pm$ defining the black hole. This differed from its usual $AdS_3$ interpretation in an essential way -- the target space isometry currents were related to quantum numbers appearing in the partition function by a boost in the former case and by a rotation in the latter.

However, the partition function includes discrete series primaries (even in the untwisted sector) corresponding to geodesics that explore the region with CTCs. This unsavoury situation could mean that the actual set of states is a subset, perhaps restricted by the $\beta<0$ condition. If so, the internal CFT will have to be chosen so that primaries with $\beta>0$ are not physical. In this context, it is instructive to examine the wavefunctions for scalar fields near the BTZ singularity $r^2=0$. It is easy to show that the relevant eigenvalue problem reduces to $(\frac{d^2}{dy^2}-\frac{4M}{J^2}\frac{d}{dy}+\frac{4}{J^4}\beta)\F=0$ (where $y=r^2$) near $r^2=0$. If $\beta<0$, this only has exponential solutions $\F_\pm(y)\propto\exp[\frac{2y}{J^2}(M\pm\sqrt{M^2-\beta})]$ which can be interpreted to mean that there is no probability current (at the level of spacetime QFTs) into or from the region $r^2<0$. One can expect that the tensor product of such exponential wavefunctions will give only exponentials. Thus, the OPE of such representations in the CFT will also close among themselves. This is perhaps satisfactory. Of course, we have to check that the level matching condition suffices to ensure that OPEs are {\em single-valued} on the worldsheet.

Alternatively, we may orbifold the spectrum read off here by the T-duality symmetry as proposed in~\cite{NS}. It will be interesting to study the {\em target space} interpretation of primaries twisted under T-duality. We also wonder if this orbifold involves a projection onto discrete energy (and angular momentum) eigenvalues, similar to those appearing in Satoh's proposal~\cite{Satoh}. This proposal was constructed by starting with states having $J_+=\pm J_-$. As we already observed, one set of such timelike geodesics \eqref{eqn:TimelikeGrazingHorizons} does not necessarily satisfy $\beta<0$ and hence, may pass into the region with CTCs. Finally, a more radical conclusion could be that the regions containing CTCs are tolerated in string theory. 

A more technical question is whether it is possible to rewrite the proposed partition function \eqref{eqn:ppsedbtzsl2} (or propose another modular invariant spectrum) in a manner that replaces $|\vartheta_1|^{-2}$ with its analytic continuation natural to $\hat{\mathfrak{sl}}(2)_k$ in the hyperbolic basis. Interestingly, one possible answer to this question arises from a double Wick rotation from $AdS_3$ to BTZ.

Consider the global parametrisation $\exp[\frac{i\s_2}{2}(t_{\text{g}}+\f_{\text{g}})]\exp(\rho\s_3)\exp[\frac{i\s_2}{2}(t_{\text{g}}-\f_{\text{g}})]$ of $AdS_3$. Here, $\f_{\text{g}}$ denotes a $2\p$-periodic angular coordinate and $t_{\text{g}}$ denotes a non-compact timelike coordinate. Now, decompactify the $\f_{\text{g}}$ direction\footnote{The $AdS_3$ metric $ds^2=-dt_{\text{g}}^2+d\rho^2+\rho^2d\f_{\text{g}}^2$ for small $\rho$ is that of plane parametrised by polar coordinates $(\rho,\f_{\text{g}})$. Hence, decompactifying $\f_{\text{g}}$ maps the single point $\rho=0$ to an entire line parametrised by $\f_{\text{g}}$.} and then perform the double Wick rotation:
\be
\begin{split}
    (\f_{\text{g}},t_{\text{g}})=(i\hat{t},i\hat{\f})\,,
\end{split}\qquad\qquad\qquad
\begin{split}
\begin{aligned}
    \hat{t}&=r_+t-r_-\f\,,\\
    \hat{\f}&=r_+\f-r_-t\,.
\end{aligned}\label{eqn:DoubleWickRotate}
\end{split}
\ee
Under this double Wick rotation, the action (hence, metric and $B$-field) of the \SL2 WZW model transforms into that of the BTZ sigma model \eqref{eqn:WZWAction}. Also, the $2\p$-periodicity of the BTZ angular coordinate $\f$ implies the compactification (equivalently, a $\mathbb{Z}$ orbifold)
\be
(\hat{\f},\hat{t})\sim(\hat{\f},\hat{t})+(2\p r_+,-2\p r_-)\,.
\ee
As a result, one may expect that the BTZ partition function is obtained by setting the temperature and chemical potential in the thermal $AdS_3$ partition function obtained by Maldacena et al.~\cite{mog2} to be $(\beta,\m\beta)=(2\p r_+,-2\p ir_-)$. This results in the following modular invariant
\be
Z_{\text{WickRotated}}=r_+\sqrt{\frac{k-2}{\tau_2}}\sum_{m,n\in\mathbb{Z}}\frac{e^{-\frac{\p k}{\tau_2}r_+^2|n\tau-m|^2+\frac{2\p}{\tau_2}\left(r_+(n\tau_1-m)-ir_-n\tau_2\right)^2}}{\vartheta_1(-i\D_-(n\tau-m)|\tau)\bar{\vartheta}_1(i\D_+(n\bar{\tau}-m)|\bar{\tau})}\,.\label{eqn:WickRotatedPartFunc}
\ee
In contrast to the partition function \eqref{eqn:ppsedbtzsl2}, notice that the denominator $\vartheta_1$ and $\bar{\vartheta}_1$ have purely imaginary arguments for $\tau\in\mathbb{R}$. This is natural to the hyperbolic basis for the following reason. Just like the global coordinates $(t_{\text{g}}+\f_{\text{g}},t_{\text{g}}-\f_{\text{g}})$, the arguments of $\vartheta_1$ and $\bar{\vartheta}_1$ are chemical potentials conjugate to zero modes $(J_{+,0}^{(0)},J_{-,0}^{(0)})$ of the elliptic generator. So, the double Wick rotation \eqref{eqn:DoubleWickRotate} maps them to chemical potentials conjugate to the hyperbolic generator. This allows us to interpret the sum in \eqref{eqn:WickRotatedPartFunc} as a sum over twists and shifts, that would naturally arise from the orbifolding  condition \eqref{eqn:orbifolding} that defines the BTZ black hole.

Exactly as in the case of thermal $AdS_3$~\cite{mog2}, the contribution $-\frac{\p k}{\tau_2}r_+^2|n\tau-m|^2$ to the exponential is the zero mode WZW action for the sector with $\f(z+2\p,\bar{z}+2\p)=\f(z,\bar{z})+2\p n$ and $\f(z+2\p\tau,\bar{z}+2\p\bar{\tau})=\f(z,\bar{z})+2\p m$. We may attempt to expand this modular invariant as a $q$-series to read off the spectrum it encodes. If $r_-$ were purely imaginary, this simply reveals the spectrum of a CFT with thermal $AdS_3$ target space having $(\beta,\m\beta)=(2\p r_+,-2\p ir_-)$. However, for purely real $r_-$, it is not clear to us how an expansion that reveals the Lorentzian BTZ spectrum can be performed.

Finally, it will of be some interest to study the spacetime modular transformation~\cite{malstrom} between Euclidean BTZ and thermal $AdS_3$ in terms of the partition functions \eqref{eqn:ppsedbtzsl2} and \eqref{eqn:WickRotatedPartFunc}.

\appendix
\section{The \texorpdfstring{\SL2}{SL(2,R)} WZW model}\label{sec:Conventions}
In this appendix, we present the details of the \SL2 WZW model to clarify our conventions. The action for this model is given by
\begin{align}     
S_{\text{WZW}}[g]&=\frac{k}{8\pi}\int d\tau \, d\sigma\,\sqrt{-h}\,
\trace\left(\del_a\,g^{-1}\del^a\,g\right) + k\, \Gamma_{\text{WZ}}[g]\,,\label{eqn:WZWAction}\\
\Gamma_{\text{WZ}}[g]&=-\frac{1}{12\pi}\int\e^{abc}\,
\trace\left(\del_ag\,g^{-1}\del_bg\,g^{-1}\del_cg\,g^{-1}\right)\,.
\end{align}
Here, the trace is calculated in the 2-dimensional representation of \SL2.

We chose the normalisation of \SL2 generators such that $\trace(\tau^a \tau^b)=\frac{1}{2}\eta^{ab}$ and where $\eta_{ab}={\text{diag}}(-1,1,1)$. More specifically, $\tau^0=\frac{i\sigma^2}{2}$, $\tau^1=\frac{\sigma^1}{2}$ and $\tau^2=\frac{\sigma^3}{2}$. The WZW currents are defined as $J_+=-k\del_+g\,g^{-1}$ and $J_-= kg^{-1}\,\del_-g$. Here, $\del_\pm$ denote derivatives with respect to worldsheet lightcone coordinates $x^\pm=\tau\pm\s$. The equations of motion that follow from \eqref{eqn:WZWAction} are $\del_\pm J_\mp=0$. The components $J^{(a)}_\pm$ of currents are defined by $J^{(a)}_\pm=\trace(\tau^a J_\pm)$ and $J^{(\pm)}_+=J^{(0)}_+\pm J^{(1)}_+$. The stress tensor (defined as
$T_{ab}=\frac{-4\pi}{\sqrt{-h}}\frac{\delta S_{\text{WZW}}}{\delta h^{ab}}$) turns out to
be $T_{\pm\pm}=\frac{1}{k}\eta_{ab}J^{(a)}_\pm J^{(b)}_\pm$ in terms of the currents. In the $+D_1^+$ charts \eqref{eqn:param}, the WZW action evaluates to 
\begin{multline}
S_{\text{WZW}}[g] =\frac{-k}{2\pi}\int dx^+ \, dx^- \,\Bigl[(r_-^2 + r_+^2-r^2) \partial_+ t \partial_- t + \frac{ r^2 \partial_+ r \partial_- r}{\left(r^2-r_-^2\right) \left(r^2-r_+^2\right)} + r^2\partial_- \f \partial_+ \f  \\ -  r_- r_+  (\partial_-t   \partial_+\f  +  \partial_-\f \partial_+t) + \left(r^2 - r_+^2 \right) ( \partial_+\f \partial_-t - \partial_+t \partial_-\f) \Bigr]\,,
\end{multline}
and the components of currents to be diagonalised upon quantisation are expressed as
\begin{subequations}\label{eqn:D1plusCurrents}
\begin{align}
J_+^{(2)}&=\frac{-k}{r_+-r_-}\biggl[\del_+\f\left(r^2-r_+r_-\right)+\del_+t\left(r_+^2+r_-^2-r_+r_--r^2\right)\biggr]\,,\\
J_-^{(2)}&=\frac{k}{r_++r_-}\biggl[\del_-\f\left(r^2+r_+r_-\right)+\del_-t\left(r^2-r_+^2-r_-^2-r_+r_-\right)\biggr]\,.
\end{align}
\end{subequations}

From the equations of motion, it follows that the currents $J_+$ and $J_-$ evaluated along a classical solution are purely left- and right- moving, respectively. Hence, they admit the following mode expansions in terms of worldsheet lightcone coordinates:
\be
J_+^{(a)}=\sum_{n\in\mathbb{Z}} J^{(a)}_{+,n}\,e^{-i n x^+}\qquad\qquad 
J_-^{(a)}=\sum_{n\in\mathbb{Z}} J^{(a)}_{-,n}\,e^{-i n x^-}
\ee
Expressing the components in terms of the coordinates $(r,t,\f)$ and their corresponding momenta (each, indexed by the worldsheet coordinate $\sigma$) allows us to compute the following equal-$\tau$ Poisson brackets
\begin{subequations}
\begin{align}
\left\{J_+^{(2)}(\cdot,\sigma),J_+^{(2)}(\cdot,\sigma^\prime)\right\}&=\pi k\partial_\sigma\delta(\sigma-
\sigma^\prime)\,,\\
\left\{J_+^{(2)}(\cdot,\sigma),J_+^{(\pm)}(\cdot,\sigma^\prime)\right\}&=\mp2\pi J_+^{(\pm)}(\cdot,\sigma^\prime)\delta(\sigma-\sigma^\prime)\,,\\
\left\{J_+^{(+)}(\cdot,\sigma),J_+^{(-)}(\cdot,\sigma^\prime)\right\}&=4\pi J_+^{(2)}(\cdot,\sigma^\prime)\delta(\sigma-\sigma^\prime)-2\pi k\partial_\sigma\delta(\sigma-\sigma^\prime)\,.
\end{align}
\end{subequations}
We may Fourier transform these Poisson brackets evaluated along a classical solution and then promote them to commutators via $[\cdot,\cdot]=i\{\cdot,\cdot\}$. This canonical quantisation results in the $\hat{\mathfrak{sl}}(2)_k$ algebra, expressed in the hyperbolic basis as
\begin{subequations}\label{eqn:HyperbolicBasis}
\begin{align}
\left[J^{(2)}_{+,n},J^{(2)}_{+,m}\right]&=\frac{k}{2}n\delta_{n+m,0}\,,\\
\left[J^{(2)}_{+,n},J^{(\pm)}_{+,m}\right]&=\mp iJ^{(\pm)}_{+,n+m}\,,\label{eqn:HyperbolicBasis:ifactor}\\
\left[J^{(+)}_{+,n},J^{(-)}_{+,m}\right]&=2iJ^{(2)}_{+,n+m}-kn\delta_{n+m,0}\,.
\end{align}
\end{subequations}
The right-moving current modes $J^{(a)}_{-,n}$ also satisfy the same algebra.

\section{Special cases} \label{sec:SpecialCases}
In this appendix, we shall consider some special cases of previous considerations, namely the non-rotating $(r_- = 0)$ and the extremal $(r_+ = r_-)$ black holes.

In the case of the non-rotating black hole, the BTZ identification becomes the axial action by $e^{\p r_+\s^3}$. At the level of geodesics, we see that $\beta= L^2r_+^2>0$. Therefore, all the timelike geodesics enter the region with closed timelike curves. We also observe that $\D_+=\D_-$ which implies $\theta_+=\theta_-$ in the identification \eqref{eqn:CurrentsIdentify} of currents. With this substitution, the interpretation of quantum numbers appearing in the partition function follows without any change. 

However, this is not the case with the extremal limit. Naive substitution of $r_-=r_+$ in the parametrisation \eqref{eqn:param} gives singular matrices. Further, substituting $r_+=r_-$ in the currents evaluated along geodesics (say \eqref{eqn:timelikecurrents}) makes them singular. So, we need to proceed with a little care.
To realise the left and right \SL2 symmetries of the extremal BTZ black hole, we need to parametrise \SL2 by charts such that the BTZ identification $\f\sim\f+2\pi$ takes the form discussed in the work of Maldacena and Strominger~\cite{malstrom}. Such a chart is
\be
e^{\frac{1}{2}(t+\f)\s^+}
\begin{pmatrix}
-\frac{r_+}{\rho} & \frac{1}{2 \rho} \\
-\rho & -\frac{\rho}{2 r_+}
\end{pmatrix}
e^{-r_+(t-\f)\s^3}\,,
\ee
where $\s^\pm=\s^1\pm i\s^2$ and $\rho^2=r^2-r_+^2$. With this parametrisation, the energy $E$ and angular momentum $L$ of a geodesic may be expressed in terms of the left parabolic generator $J_+^{(+)}=\frac{1}{2}\trace(\s^+J_+)$ and the right hyperbolic generator $J_-^{(2)}$ evaluated along the geodesic as
\be
J_+^{(+)}=-\frac{k}{2}(E+L)\,,\qquad\qquad J_-^{(2)}=-\frac{k}{2}\frac{E-L}{2r_+}\,.
\ee
Twisted sectors of the extremal case are also different from those of generic BTZ black holes. The analogue of twisting \eqref{eqn:specflowdefn} in the extremal case is
\be
g_\Omega(\tau,\s)=e^{\frac{\Omega}{2}x^+\s^+}\tilde{g}(\tau,\s)e^{-\Omega r_+x^-\s^3}\,,\qquad\qquad x^\pm=\tau\pm\s\,,
\ee
and components of the left current evaluated along the twisted solution are related to those before twisting in a manner different from the $r_+\neq r_-$ case. In particular, the component of the left current to be diagonalised upon quantisation retains its value after twisting
\begin{subequations}\label{eqn:TwistedCurrentsExtremal}
\begin{align}
    J_+^{(+)}&=\tilde{J}_+^{(+)}\,, &&& J_-^{(+)}&=e^{-2r_+x^-\Omega}\tilde{J}_-^{(+)}\,,\\
    J_+^{(-)}&=\tilde{J}_+^{(-)}+k\Omega+2\Omega x^+\tilde{J}_+^{(2)}+(\Omega x^+)^2\tilde{J}_+^{(+)}\,, &&&
    J_-^{(-)}&=e^{+2r_+x^-\Omega}\tilde{J}_-^{(-)}\,,\label{eqn:LeftJMinusTwist}\\
    J_+^{(2)}&=\tilde{J}_+^{(2)}+\Omega x^+\tilde{J}_+^{(+)}\,, &&&
    J_-^{(2)}&=\tilde{J}_-^{(2)}-kr_+\Omega\,.
\end{align}
\end{subequations}
We may verify that the currents evaluated along the twisted solution also satisfy the same Poisson brackets \eqref{eqn:HyperbolicBasis} as those evaluated along the solution before twisting.

The relation between values of the stress tensor evaluated along twisted and untwisted solutions follows from \eqref{eqn:TwistedCurrentsExtremal}; it is 
\be
T_{++}=\tilde{T}_{++}-\Omega\tilde{J}_+^{(+)}\,,\qquad\qquad T_{--}=\tilde{T}_{--}-2r_+\Omega\tilde{J}_-^{(2)}+kr_+^2\Omega^2\,.
\ee
This agrees with $L_0\pm\bar{L}_0$ read off from the expansion \eqref{eqn:PartFuncExpand}, if we identify eigenvalues $(J_+,J_-)$ and $(\tilde{J}_+,\tilde{J}_-)$ of $(J_+^{(+)},J_-^{(2)})$ after and before spectral flow respectively as
\be
\begin{pmatrix}
\tilde{J}_+\\J_+\\\tilde{J}_-
\\
J_-
\end{pmatrix}=\frac{k}{2}\begin{pmatrix}
\frac{1}{2}&-\frac{1}{2}&0&0\\
\frac{1}{2}&-\frac{1}{2}&0&0\\
0&0&-\cosh\q_-&\sinh\q_-\\
0&0&-\sinh\q_-&\cosh\q_-
\end{pmatrix}\begin{pmatrix}
w'+t_1-\frac{n'}{k}\\
w-t_1+\frac{n}{k}\\
w'+t_1+\frac{n'}{k}\\
w-t_1-\frac{n}{k}
\end{pmatrix}\,,\label{eqn:CurrentsIdentifyExtremal}
\ee
where $e^{-\q_-}=\D_+=2r_+$. Notice that the matrix used has zero determinant. In this sense, it is different from the generic BTZ case \eqref{eqn:CurrentsIdentify} as well as the $AdS_3$ case \eqref{eqn:CurrentsIdentifyAdS3}. Exactly as in those cases, we can verify that with this identification, the spectral flow parameter is $\Omega=-(w+w')$. The energy and angular momentum before and after spectral flow have the same expressions \eqref{eqn:PreTwistIdentify} and \eqref{eqn:PostTwsistIdentify} as the generic case, but with $M=2r_+^2=J$. Hence, this identification is consistent with angular momentum quantisation and allows the energy to have an arbitrary real value. It is also quantitatively consistent with the Noether ambiguity \eqref{eqn:TranslationOperatorTwist} discussed earlier.

\section{Free boson conformal blocks and the \texorpdfstring{$\vartheta_1$}{vartheta1}-function} \label{sec:Notation}
The free boson conformal blocks appearing in the partition function \eqref{eqn:ppsedbtzsl2} are defined as
\be 
\z\oao{w}{m}(k) =\sqrt{ \frac{k}{\tau_2}}\exp\left(-\frac{\p k}{\tau_2}|w\tau-m|^2\right)\,.\label{eqn:zetadef}
\ee 
The $\vartheta_1$-function is defined as
\be
\vartheta_1(v|\tau)=\sum_{p \in \mathbb{Z}}e^{\p i\tau(p+\frac{1}{2})^2+2\p i(v+\frac{1}{2})(p+\frac{1}{2})}\,.\label{eqn:vartheta1sumrep} 
\ee
It can also be expressed (letting, $q=e^{2\p i\tau}$) as the infinite product
\be
\vartheta_1(v|\tau)=-2q^{1/8}\sin\p v\prod_{p=1}^{\infty}(1-e^{2i\p v}q^p)(1-q^p)(1-e^{-2i\p v}q^p)\,.\label{eqn:vartheta1prodrep}
\ee
To expand $|\vartheta_1|^{-2}$ as a power series to obtain \eqref{eqn:sl2expansion}, we start with the product representation
\begin{multline}
\frac{1}{|\vartheta(v|\tau)|^2}=\frac{(q\bar{q})^{-1/8}}{4\sin(\p v)\sin(\p\bar{v})}\prod_{p=1}^{\infty}\frac{1}{(1-q^p)(1-\bar{q}^p)}\\
\times\frac{1}{(1-e^{2iv\p}q^p)(1-e^{-2iv\p}q^p)(1-e^{-2i\bar{v}\p}\bar{q}^p)(1-e^{2i\bar{v}\p}\bar{q}^p)}\,.\label{eqn:rc1.1}
\end{multline}
Each factor in the product may be expanded into a geometric series. However, for the series to converge, a different expansion needs to performed depending on ${\text{Im}}(v)$. In the partition function \eqref{eqn:ppsedbtzsl2}, we have $v=s_1\tau-s_2$ with $s_1,s_2\in(0,1)$ and hence $0<{\text{Im}}(v)<\tau_2$. A convergent series expansion of the prefactor from \eqref{eqn:rc1.1} in this domain is
\be
\begin{aligned}
\frac{1}{4\sin(\p v)\sin(\p\bar{v})}&=\frac{e^{-2\p s_1\tau_2}}{(1-e^{2i\p(s_1\tau_1-s_2)}e^{-2\p s_1\tau_2})(1-e^{-2i\p(s_1\tau_1-s_2)}e^{-2\p s_1\tau_2})}\\
&=e^{-2\p s_1\tau_2}\sum_{P_{+,0},P_{-,0}=0}^\infty e^{2i\p(s_1\tau_1-s_2)(P_{+,0}-P_{-,0})}e^{-2\p s_1\tau_2(P_{+,0}+P_{-,0})}\,.
\end{aligned}
\ee 
Its convergence follows from $|e^{\pm2i\p(s_1\tau_1-s_2)}e^{-2\p s_1\tau_2}|<1$ (for $\tau_2>0$). A similar expansion can be used for each factor in the infinite product part of \eqref{eqn:rc1.1}. These expansions will converge because $s_1<p\in\{1,2,3,\ldots\}$. The resulting expansion is
\be
\frac{1}{|\vartheta(s_1\tau-s_2|\tau)|^2}=\sum_{q,\bar{q},N,\bar{N}}e^{2\p is_2(\bar{q}-q)+2\p i\tau_1[s_1(q-\bar{q})+N-\bar{N}]-2\p\tau_2[s_1(1+q+\bar{q})-\frac{1}{4}+N+\bar{N}]}\,,
\ee 
where the sum $\sum_{q,\bar{q},N,\bar{N}}$ is shorthand for a sum over $P^+_{\pm,p}$, $P^-_{\pm,p}$ (with $p\in\{1,2,3,\ldots\}$) and $P_{\pm,p}$ (with $p\in\{0,1,2,\ldots\}$), each ranging over $\{0,1,2,\ldots\}$ with $(q,\bar{q})$ and $(N,\bar{N})$ defined as
\begin{subequations}\label{eqn:degeneracymatch}
\begin{align}
q&=P_{+,0}+\sum_{p=1}^{\infty}(P^+_{+,p}-P^-_{+,p})\,,&\bar{q}&=P_{-,0}+\sum_{p=1}^{\infty}(P^+_{-,p}-P^-_{-,p})\,,\label{eqn:degeneracymatch:qbarq}\\
N&=\sum_{p=1}^{\infty}p(P_{+,p}+P^+_{+,p}+P^-_{+,p})\,,&\bar{N}&=\sum_{p=1}^{\infty}p(P_{-,p}+P^+_{-,p}+P^-_{-,p})\,.
\end{align}
\end{subequations}

\acknowledgments
O.N.\ and A.S.\ thank IISER Mohali for providing hostel accommodation for part of their MS~thesis work. They were supported, in part, through the INSPIRE~SHE. All of us thank Jitsi and Overleaf for providing valuable and free services which enabled us to engage in a fruitful collaboration.

\paragraph{Note added.}
As this manuscript was being finalised, the paper~\cite{ST} appeared on the arXiv where the modular invariant \eqref{eqn:WickRotatedPartFunc} was presented as the partition function of the Euclidean BTZ black hole.


\begin{thebibliography}{99}

\bibitem{btz}
M.~Banados, M.~Henneaux, C.~Teitelboim and J.~Zanelli, \emph{Geometry of the (2+1) black hole}, \emph{Phys. Rev. D} {\bf 48} (1993) 1506.
Erratum: [Phys. Rev. D {\bf 88}, 069902 (2013)]
doi:10.1103/PhysRevD.48.1506, 10.1103/PhysRevD.88.069902
[gr-qc/9302012].

\bibitem{mog}
J.~M.~Maldacena and H.~Ooguri, \emph{Strings in AdS(3) and SL(2,R) WZW model 1.: The Spectrum}, \emph{J. Math. Phys.} {\bf 42}, (2001) 2929. doi:10.1063/1.1377273 [hep-th/0001053].

\bibitem{mog2}
J.~M.~Maldacena, H.~Ooguri and J.~Son, \emph{Strings in AdS(3) and the SL(2,R) WZW model. Part 2. Euclidean black hole}, \emph{J. Math. Phys.} {\bf 42} (2001) 2961. doi:10.1063/1.1377039 [hep-th/0005183].

\bibitem{kpy}
K.~P.~Yogendran, \emph{Closed Strings in the 2D Lorentzian Black Hole}, arXiv:1808.10109 [hep-th].

\bibitem{Israel}
D.~Israel, C.~Kounnas and M.~P.~Petropoulos, \emph{Superstrings on NS5 backgrounds, deformed AdS(3) and holography}, \emph{JHEP} {\bf 10} (2003) 028. doi:10.1088/1126-6708/2003/10/028 [arXiv:hep-th/0306053 [hep-th]].

\bibitem{Mukunda}
J.~G.~Kuriyan, N.~Mukunda, E.~C.~G.~Sudarshan, \emph{Master Analytic Representation: Reduction of $O(2,1)$ in an $O(1,1)$ Basis}, \emph{J. Math. Phys.} {\bf 9}, (1968) 2100.
doi: 10.1063/1.1664551

\bibitem{NS}
M.~Natsuume and Y.~Satoh, \emph{String theory on three-dimensional black holes}, \emph{Int. J. Mod. Phys. A} {\bf 13} (1998) 1229. doi:10.1142/S0217751X98000585 [hep-th/9611041].

\bibitem{kraus}
S.~Hemming, E.~Keski-Vakkuri and P.~Kraus, \emph{Strings in the extended BTZ space-time}, \emph{JHEP} {\bf 0210} (2002) 006. doi:10.1088/1126-6708/2002/10/006 [hep-th/0208003].

\bibitem{malstrom}
J.~M.~Maldacena and A.~Strominger, \emph{AdS(3) black holes and a stringy exclusion principle}, \emph{JHEP} {\bf 9812}, (1998) 005. doi:10.1088/1126-6708/1998/12/005 [hep-th/9804085].

\bibitem{Esko}
S.~Hemming and E.~Keski-Vakkuri, \emph{The Spectrum of strings on BTZ black holes and spectral flow in the SL(2,R) WZW model}, \emph{Nucl. Phys. B} {\bf 626}, (2002) 363. doi:10.1016/S0550-3213(02)00021-4 [hep-th/0110252].

\bibitem{RR}
M.~Rangamani and S.~F.~Ross, \emph{Winding tachyons in BTZ}, \emph{Phys. Rev. D} {\bf 77} (2008) 026010. doi:10.1103/PhysRevD.77.026010 [arXiv:0706.0663 [hep-th]].

\bibitem{GPS}
S.~B.~Giddings, J.~Polchinski and A.~Strominger, \emph{Four-dimensional black holes in string theory}, \emph{Phys. Rev. D} {\bf 48} (1993) 5784-5797. doi:10.1103/PhysRevD.48.5784 [arXiv:hep-th/9305083 [hep-th]].

\bibitem{Cruz}
N.~Cruz, C.~Martinez and L.~Pena, \emph{Geodesic structure of the (2+1) black hole}, \emph{Class. Quant. Grav.} {\bf 11} (1994) 2731-2740. doi:10.1088/0264-9381/11/11/014 [arXiv:gr-qc/9401025 [gr-qc]].

\bibitem{Troost1}
J.~Troost, \emph{Winding strings and AdS(3) black holes}, \emph{JHEP} {\bf 0209} (2002) 041. doi:10.1088/1126-6708/2002/09/041 [hep-th/0206118].

\bibitem{Ooguri}
J.~de Boer, H.~Ooguri, H.~Robins and J.~Tannenhauser, \emph{String theory on AdS(3)}, \emph{JHEP} {\bf 12} (1998) 026.
doi:10.1088/1126-6708/1998/12/026
[arXiv:hep-th/9812046 [hep-th]].

\bibitem{Eberhardt}
L.~Eberhardt, \emph{Partition functions of the tensionless string}, \emph{JHEP} {\bf 03} (2021), 176.
doi:10.1007/JHEP03(2021)176
[arXiv:2008.07533 [hep-th]].

\bibitem{Satoh}
Y.~Satoh, \emph{Ghost - free and modular invariant spectra of a string in SL(2,R) and three-dimensional black hole geometry}, \emph{Nucl. Phys. B} {\bf 513} (1998), 213-228.
doi:10.1016/S0550-3213(97)00701-3
[arXiv:hep-th/9705208 [hep-th]].

\bibitem{Hemming}
S.~Hemming, \emph{On free field realizations of strings in BTZ}, \emph{Int. J. Mod. Phys. A} {\bf 19} (2004), 1579-1588.
doi:10.1142/S0217751X04017859
[arXiv:hep-th/0304009 [hep-th]].

\bibitem{vijay}
V.~Balasubramanian, A.~Kar and G.~S\'arosi, \emph{Holographic Probes of Inner Horizons}, \emph{JHEP} {\bf 06} (2020), 054.
doi:10.1007/JHEP06(2020)054
[arXiv:1911.12413 [hep-th]].

\bibitem{HongLiu}
G.~Festuccia and H.~Liu, \emph{Excursions beyond the horizon: Black hole singularities in Yang-Mills theories. I.}, \emph{JHEP} {\bf 04} (2006), 044.
doi:10.1088/1126-6708/2006/04/044
[arXiv:hep-th/0506202 [hep-th]].

\bibitem{ST}
S.~K.~Ashok and J.~Troost, \emph{Twisted Strings in Three-dimensional Black Holes},
arXiv:2112.08784.




\end{thebibliography}
\end{document}